\documentclass{ieeeaccess}
\usepackage{amsmath,amssymb,amsfonts,arydshln}
\usepackage{algorithmic}
\usepackage{graphicx}
\usepackage{textcomp}
\usepackage[numbers,sort&compress]{natbib}

\usepackage{url}

\newcommand{\nbh}[1]{\texttt{#1}}
\usepackage{booktabs}

\def\BibTeX{{\rm B\kern-.05em{\sc i\kern-.025em b}\kern-.08em  T\kern-.1667em\lower.7ex\hbox{E}\kern-.125emX}}

\begin{document}
\history{Date of publication xxxx 00, 0000, date of current version xxxx 00, 0000.}
\doi{00.0000/ACCESS.2020.DOI}

\title{nnAudio: An on-the-fly GPU Audio to Spectrogram Conversion Toolbox Using 1D Convolution Neural Networks}
\author{\uppercase{Kin Wai Cheuk}\authorrefmark{1,2}, \IEEEmembership{Student Member, IEEE},
\uppercase{Hans Anderson\authorrefmark{3}, Kat Agres\authorrefmark{2} \IEEEmembership{Member, IEEE}, Dorien Herremans\authorrefmark{1,2}}, \IEEEmembership{Senior Member, IEEE},}
\address[1]{Information Systems, Technology and Design, Singapore University of Technology and Design, 8 Somapah Rd, Singapore 487372 (e-mail: kinwai\_cheuk@mymail.sutd.edu.sg, dorien\_herremans@sutd.edu.sg)}
\address[2]{Institute of High Performance Computing, Agency for Science, Technology and Research, 1 Fusionopolis Way, \#16-16 Connexis, Singapore 138632 (e-mail: kat\_agres@ihpc.a-star.edu.sg)}
\address[3]{Blue Mangoo Software (email: hans@bluemangoo.com)}
\tfootnote{This work is supported by the SUTD-MIT IDC grant IDG31800103, MOE Grant no. MOE2018-T2-2-161, and SING-2018-02-0204. }

\markboth
{Cheuk \headeretal: Preparation of Papers for IEEE TRANSACTIONS and JOURNALS}
{Cheuk \headeretal: Preparation of Papers for IEEE TRANSACTIONS and JOURNALS}

\corresp{Corresponding author: Dorien Herremans (e-mail: dorien\_herremans@sutd.edu.sg).}

\begin{abstract}
Despite recent developments in neural network models that use raw audio as input, state-of-the-art results from tasks such as automatic music transcription (AMT) and automatic speech recognition (ASR) are often still achieved by using frequency domain features such as spectrograms as input. Converting time domain waveforms to frequency domain spectrograms is typically considered to be a prepossessing step done before model training. This approach, however, has several drawbacks. First, it takes a lot of hard disk space to store different frequency domain representations. This is especially true during the model development and tuning process, when exploring various types of spectrograms for optimal performance. Second, if another dataset is used, one must process all the audio clips again before the network can be retrained. In this paper, we integrate the time domain to frequency domain conversion as part of the model structure, and propose a neural network based toolbox, \nbh{nnAudio}, which leverages 1D convolutional neural networks to perform time domain to frequency domain conversion during feed-forward. It allows on-the-fly spectrogram generation without the need to store any spectrograms on the disk. This approach also allows back-propagation on the waveforms-to-spectrograms transformation layer, which implies that this transformation process can be made trainable, and hence further optimized by gradient descent. nnAudio reduces the waveforms-to-spectrograms conversion time for 1,770 waveforms (from the MAPS dataset) from $10.64$ seconds with librosa to only $0.001$ seconds for Short-Time Fourier Transform (STFT), $18.3$ seconds to $0.015$ seconds for Mel spectrogram, $103.4$ seconds to $0.258$ for constant-Q transform (CQT), when using GPU on our DGX work station with CPU: Intel(R) Xeon(R) CPU E5-2698 v4 @ 2.20GHz Tesla v100 32Gb GPUs. (Only 1 GPU is being used for all the experiments.) We also further optimize the existing CQT algorithm, so that the CQT spectrogram can be obtained without aliasing in a much faster computation time (from $0.258$ seconds to only $0.001$ seconds).
\end{abstract}

\begin{keywords}
Convolution,
Discrete Fourier transform,
Short time Fourier transform,
Spectrogram,
CQT,
Mel Spectrogram,
Signal processing, 
Library, 
PyTorch,
GPU
\end{keywords}

\titlepgskip=-15pt

\maketitle

\section{Introduction}
\label{sec:Introduction}
\PARstart{M}{achine} learning for audio is a challenging task as raw audio waveforms are time domain features: Because only the amplitudes over time can be observed directly from a time domain signal, a model with a long-term memory is required to retrieve meaningful information from only the amplitudes. It is therefore necessary to convert these features into the frequency domain in order to get the implicit frequency information from the waveforms. Despite recent advances in time domain based models, such as WaveNet~\cite{Oord2016WaveNetAG}, many models using frequency domain features achieve better results than models using time domain features~\cite{hertel2016comparing,Pons2017EndtoendLF}. WaveNet is a model designed to take raw waveforms as input, and has inspired several recent audio related machine learning models~\cite{Juvela2019GlotNetARW,Rushe2019AnomalyDI,Yoshimura2019SpeakerdependentWD}. Despite these advances, countless models are still using frequency domain features as the model's input for various tasks due to their superior performance~\cite{choi2019zero, gururani2019attention,doras2019cover, doras2019use, meseguer2019conditioned,gillickestimating,balke2019learning,wigginsguitar,nidadavolu2019investigation,Neumann2019ImprovingSE,alAbdullah2019ArtificialNN,Kelz2019TowardsIP,hantrakulfast,taenzerinvestigating,park2019bi,kim2019adversarial,leeautomatic,fuhierarchical,wang2019adaptive}. Therefore, there is still value in developing a faster time-frequency conversion computation method, which is what we propose in this paper. 

The Discrete Fourier transform is one of the classic algorithms that allows conversion from the time domain into the frequency domain, but other algorithms may be used depending on the task. For example, the constant-Q transform (CQT) may be more suitable for processing music audio clips~\cite{brown1991calculation}. In order to find the audio transformation methods best suited to a specific task, trial and error may be needed. The naive way to achieve this is to convert audio clips to different frequency domain representations, and save each of the representations on the hard disk. After that, the neural networks are trained using these different representations, and the best performing model is selected. Even when the best representation is known, the transformation parameters such as window size and number of frequency bins can be further fine-tuned to obtain a better result. There are two problems with this naive approach. First, a considerable amount of hard disk space is required to store different frequency domain representations. Second, if another dataset is used, one must process all the audio clips before the actual neural network can be trained. In this paper, we propose a neural network based audio processing toolbox, \nbh{nnAudio}, which allows a GPU to generate on-the-fly spectrograms. Our spectrogram generation approach is based on trainable neural networks.

Using a GPU to process audio signals is not a new concept. \nbh{Tensorflow}~\cite{Tensorflow2015-whitepaper} has a \nbh{tf.signal} package that performs the Fast Fourier Transform (FFT) and Short-Time Fourier Transform (STFT) on GPUs. There is a high-level API, called \nbh{Keras}, for people who want to quickly build a neural network without having to work with \nbh{Tensorflow} sessions. (Tensorflow 1.14 is the prominent version at the time of this writing.) \nbh{Kapre} \cite{choi2017kapre} is the \nbh{Keras} version for GPU audio processing. Along similar lines, \nbh{PyTorch}~\cite{paszke2017automatic} has recently developed \nbh{torchaudio}, but this tool has not been fully integrated into \nbh{PyTorch} at the time of writing this paper. Furthermore, torchaudio requires Libsox as an extra dependency, and the installation often requires significant trouble-shooting~\cite{torchaudio_mac}; for example, \nbh{torchaudio} is currently not compatible with Windows 10~\cite{torchaudio_win}. Among the three tools, only Kapre and \nbh{torchaudio} support audio to Mel spectrogram conversions, and Kapre is the only implementation that supports log-frequency spectrograms and trainable kernels for time domain to frequency domain transformations, due to the neural network implementation. These, however, cannot be integrated with the popular machine learning library \nbh{PyTorch}. Despite the GPU support, \nbh{torchaudio} and \nbh{tf.signal} are not neural network based, meaning that they cannot be integrated into the neural network and perform gradient descent together. Therefore, we developed \nbh{nnAudio}~\cite{nnAudio} using \nbh{PyTorch}. More specifically, we use the 1D convolution layer to facilitate integration with other neural network layers, which makes \nbh{nnAudio} unique. This library is useful when exploring different input representation for neural network models~\cite{balamurali2019toward,kelz2016potential}.
The library is available online\footnote{\label{nnAudio} Via PyPI (nnAudio), and \url{https://github.com/KinWaiCheuk/nnAudio}}.

In the following subsections, we will briefly summarize the mathematics of the Discrete Fourier Transform (DFT). We will then discuss how to initialize a neural network to perform the STFT, Mel spectrogram and constant-Q transform (CQT) in Section-\ref{sec: PyTorch}. In section~\ref{sec: Result}, we compare speed and output of nnAudio versus a popular python signal processing library, librosa. Finally, we end with potential applications of our proposed library. 
The main contributions of this paper are (1) implementing CQT using a 1D convolutional neural network; (2) improving the existing CQT algorithm to reduce the number of steps required for the final result; and (3) providing an example to show how trainable STFT or Mel spectrograms can improve the overall model performance for a signal frequency prediction task. This paper can also serve as a guide on how to implement the DFT and its variants on other automatic differentiation libraries available in the future.

\section{Signal processing: A Quick Overview}
In this section, we will go over the basic transformation methods (DFT) used to covert signals from the time domain to the frequency domain. 

\subsection{Discrete Fourier Transform (DFT)} \label{chap:DFT}
When recording audio using any computer or mobile device, the analogue signal is converted to a digital signal before storing the data. Therefore, the audio waveforms consist of discrete data points. The Discrete Fourier Transform can be used to convert this discrete signal from the time domain to the frequency domain. Equation~\eqref{DFT2} shows the mathematical expression for the discrete Fourier transform~\cite{wang1984fast}, where $X[k]$ is the output in the frequency domain; and $x[n]$ is the $n^{th}$ sample of the audio input in the time domain. For real-valued inputs, the frequency domain output $X[k]$ for $k \in [1,N/2]$ is equal to the output $X[k]$ for $k \in [N/2,N-1]$ in reverse order, where $N$ is the window length (which is usually a power of two such as 1,024 and 2,048). To discard this redundant frequency domain information, only the first half of the frequency bins in the frequency domain will be extracted, i.e. $k \in [0,\frac{N}{2}]$. We define the DFT as a split-sum of real and complex components:



\begin{equation} 
X[k] = \sum_{n=0}^{N-1}x[n] \cos(2\pi k \frac{n}{N}) - i \sum_{n=0}^{N-1}x[n]\sin(2\pi k \frac{n}{N})
\label{DFT2}.
\end{equation}

When we use \eqref{DFT2} to compute the DFT with a 1D convolutional neural network, we can calculate the real and complex terms separately using real-valued arithmetic.

The frequency $k$ in the DFT is given in terms of normalized frequency (equivalent to cycles per window). The formula to convert the normalized frequency $k$ to the frequency $f$ in units of Hertz (Hz) is given by \eqref{f-k-conversion},

\begin{equation}
 f= k\frac{s}{N} \label{f-k-conversion},
\end{equation}

where $s$ is the sample rate and $N$ is the FFT window length.

\subsection{DFT for arbitrary frequency ranges}
Since $k$ is an integer ranging from zero to half of the window length, the DFT is only capable of resolving a finite number of distinct frequencies. For example, if the sampling rate is 44,100Hz, and the window length is 2,048, then the normalized frequencies for the DFT kernel are $k=[0,1,2,$\ldots$,1024]$ which corresponds to a DFT kernel with frequencies  $f=[0,21.53,43.07,$\ldots$, 22050]$~Hz (using \eqref{f-k-conversion}).  The frequency resolution under this setting is 21.53Hz. For comparison, the lowest two notes on a piano keyboard are $A0=27.5$ Hz and $A\#0=29.14$ Hz. With a difference of less than 2 Hz between them, the DFT of 1,024 frequency bins is not sufficient to resolve the correct note. The frequency resolution $\Delta f$ is given by \eqref{freq-resolution}. This resolution can be improved by increasing the window size $N$, however, increasing the window size results in a decrease in time resolution $\Delta t$, as shown in \eqref{time-resolution}. Therefore, we are forced to make a compromise between time and frequency resolution as per \eqref{freq-time-resolution}.

\begin{equation}
 \Delta f = \frac{s/2}{N/2} \label{freq-resolution}
\end{equation}
\begin{equation}
 \Delta t = \frac{N}{s} \label{time-resolution}
\end{equation}
\begin{equation}
 \Delta f\Delta t = 1 \label{freq-time-resolution}
\end{equation}

The vectors of the DFT transformation matrix are a basis for the set of all complex vectors of length N. This implies that applying the DFT followed by the inverse-DFT results in a perfect reconstruction of the original signal. Invertibility is important for many signal processing applications, but in information retrieval applications such as speech recognition and sound classification, it is not always necessary to use an invertible time-frequency transformation. In such cases we may want to modify the the DFT in ways that no longer result in an orthogonal set of basis vectors.

One way to modify the DFT is to change the frequencies of the basis vectors to increase or decrease the number of bins in certain parts of the spectrum. To achieve linear-scale frequency with non-integer multiples of $s/N$ in equation \eqref{f-k-conversion} we can replace $k$ with $\sigma(k) = Ak + B$, where $A$ and $B$ are two constants. To find $A$ and $B$, let $f_e$ and $f_s$ be the ending and starting frequencies of the range we want to analyse, and apply \eqref{f-k-conversion} to get \eqref{linear-frequency-DFT}, where $\mu \in \left[0, \frac{N}{2}+1 \right]$ is the number of bins chosen to be displayed in the spectrogram.

\begin{equation}
 \sigma(k) = \frac{(f_e-f_s)N}{\mu s}k + \frac{f_s N}{s}\label{linear-frequency-DFT}
\end{equation}

By the same token, we can generate basis vectors for a log-frequency spectrogram by using $\sigma(k) = Be^{Ak}$, resulting in $A=\frac{f_sN}{s}$ and $B=\frac{\ln{\frac{f_e}{f_s}}}{\mu}$ as shown in \eqref{log-frequency-DFT} below.

\begin{equation}
 \sigma(k) = \frac{f_s N}{s}\left(\frac{f_e}{f_s}\right)^\frac{k}{\mu} \label{log-frequency-DFT}
\end{equation}

Note that we use the word "basis" informally here. These formulae do not guarantee a linearly-independent set of vectors, so the basis we get from this method may in fact be rank-deficient. When using \eqref{log-frequency-DFT} or \eqref{linear-frequency-DFT}, \eqref{DFT2} becomes \eqref{DFT3}. This more general time-frequency transform permits us to focus the resolution of our spectrogram in the frequency range where it is most needed. For example, if our starting frequency is $f_s=50$Hz and the ending frequency is $f_e=6000$Hz, the linear frequency DFT kernel would have basis vectors with normalized frequency $\sigma(k \in [0,1024]) = [2.32,2.59,2.86, ..., 278.10,278.36]$. This corresponds to the frequency $f=[50,55.8,61.6,..., 5988,5994]$Hz. The frequency resolution has improved from $21.53$Hz to $5.8$Hz without changing the transform window size.

\begin{equation} X[k] = \sum_{n=0}^{N-1}x[n] \cos(2\pi \sigma(k) \frac{n}{N}) - i \sum_{n=0}^{N-1}x[n]\sin(2\pi \sigma(k) \frac{n}{N})\label{DFT3}\end{equation}

Note that this method only changes the spacing between the centres of adjacent frequency bins without affecting the width of the bins themselves. Because each bin represents a range of frequencies in a fixed-width region centred around $f$ as given in \eqref{f-k-conversion}, we will lose information if we space the bins too far apart.

In the next section, we explain how the DFT in \eqref{DFT2} and the variable-resolution DFT in \eqref{DFT3} is used to implement the short-time Fourier transform (STFT) using a convolutional neural network. The frequency scaling factor will be integrated as one of the input features in our neural network based implementation.

\section{Neural Network Based Implementations} \label{sec: PyTorch}
In this section, we will discuss how to implement the short-time Fourier transform (STFT), Mel spectrogram, and constant-Q transform (CQT) using a 1D convolutional neural network. These are then  implemented as a library (nnAudio) in PyTorch\footnotemark[\ref{nnAudio}]). The STFT is the fundamental operation for both Mel spectrogram calculation and CQT. To convert the STFT spectrogram to a Mel spectrogram we simply multiply by a Mel filter bank kernel. Similarly, the computation of the CQT also begins with the STFT, followed by multiplication with a CQT kernel. We begin this section by explaining how we use a convolutional neural network to compute the STFT.

\subsection{Short-Time Fourier Transform (STFT)} \label{subsec:STFT}
The Short-time Fourier Transform (STFT), also called the sliding-window DFT, refers to an application of the DFT wherein the signal is cut into short windows before performing the transform, rather than performing one large transform on the entire signal~\cite{Nawab:1987:SFT:42739.42745}. For audio analysis applications, this is the standard way to apply the DFT.

The STFT is usually calculated using the Cooley-tukey Fast Fourier Transform algorithm (FFT), which is preferred because it computes the DFT in $O(N \log N)$ operations, as opposed to $O(N^2)$ for the canonical DFT implementation. However, implementations of the $O(N^2)$ DFT often out-perform the $O(N \log N)$ FFT for small values of $N$ when the underlying platform supports fast vector multiplication. This is especially true when the computation is done in parallel on a GPU. Since neural network libraries typically include fast GPU-optimised convolution functions, we can compute the canonical DFT quickly on those platforms by expressing the vector multiplication in the DFT as a one-dimensional linear convolution operation. 

Discrete linear convolution of a kernel $h$ with a signal $x$ is defined as follows,
\begin{equation}
    (h * x)[n] = \sum^{M-1}_{m=0}x[n-m]h[m],\label{convolutionDefinition}
\end{equation}
where $M$ is the length of the kernel $h$. PyTorch defines a convolution function with a stride argument. The one dimensional convolution of $x$ with $h$ using a stride setting of $k$, denoted by the symbol $*^k$ is,
\begin{equation}
    (h *^k x)[n] = \sum^{M-1}_{m=0}x[k n-m]h[m].\label{convolutionWithStride}
\end{equation}
We can use convolution with stride to make fast GPU-based implementations of the short time Fourier transform (STFT). To do this, we take each basis vector of the DFT as the filter kernel, $h$ and compute the convolution with the input signal $x$ once for each basis vector. We set the stride value according to the amount of overlap we want to have between each DFT window. For example, for zero overlap, we set the stride to $N$, the length of the DFT; and for 1/2 window overlap, we set the stride to $N/2$.

Note that due to the way convolution, as defined in \eqref{convolutionDefinition} and \eqref{convolutionWithStride}, computes array indices, we need to reverse the order of elements in the DFT basis vectors when creating the convolution kernels. The following expressions are the pair of convolution kernels ($h_{\text{re}}[k,n]$ and $h_{\text{i }}[k,n]$) that represent the real and imaginary components of the $k^{th}$ DFT basis vector respectively,

\begin{align}
    h_{\text{re}}[k,n] &= \cos(2\pi k \frac{N-n-1}{N}),\\
    h_{\text{im}}[k,n] &= \sin(2\pi k \frac{N-n-1}{N}).
\end{align}
The DFT is usually computed with a function that fades the samples at the edges of each window smoothly down to near zero to avoid the high-frequency artefacts that are introduced by cutting the window abruptly at the edges~\cite{Oppenheim1989DiscretetimeSP}. Typical examples of DFT window functions include Hann, Hamming, and Blackman types. In a GPU-based DFT implementation using a convolution function with stride \eqref{convolutionWithStride}, we can implement the window smoothing efficiently by multiplying these window function elementwise with the filter kernels $h_i$ and $h_r$ before doing the convolution.

When calculating spectrograms, we typically use the Discrete Fourier Transform of length $N=2048$ or $N=4096$, but other values of $N$ are possible. We often cut the DFT windows so that they overlap each other by some amount in order to improve the time resolution. In a signal with $T$ windows, we let $X_t$ denote the DFT of the window at index $t \in [0,T-1]$. The time domain representation of the window at index $t$ will be denoted by $x_t$.




\Figure[t!](topskip=0pt, botskip=0pt, midskip=0pt)[width=\linewidth]{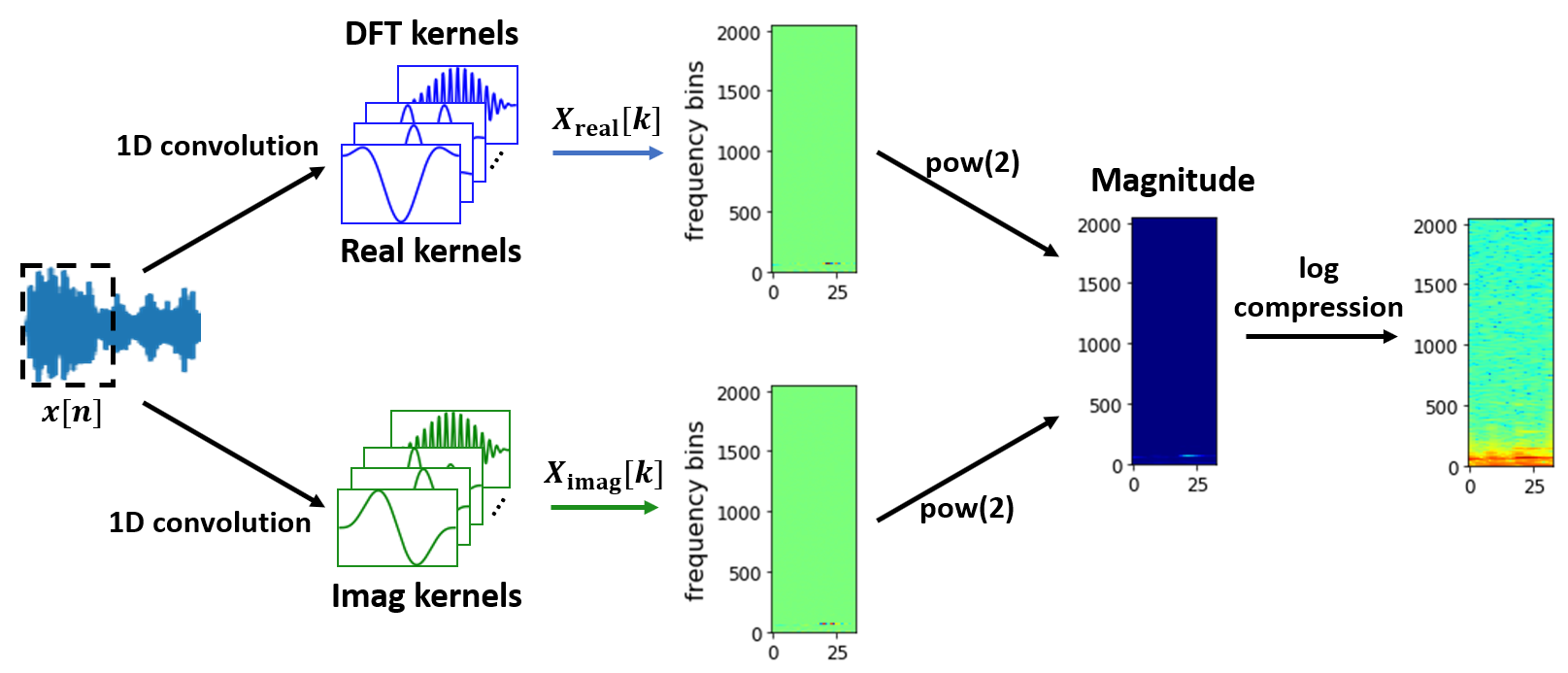}
{An STFT with a sliding window can be achieved by implementing DFT and initializing the 1D convolution kernels as cosine and sine in \nbh{PyTorch}. Applying logarithmic compression on the magnitude allows for a better visualization of the spectrogram. \label{fig: conv1d}}


Figure \ref{fig: conv1d} shows the schematic diagram for neural network based STFT. There are two most prominent advantages of implementing the STFT using a \nbh{PyTorch} 1D convolutional neural network. Firstly, it supports batch processing. Using a GPU based neural network approach, we can convert a tensor of audio clips to a tensor of spectrograms with tensor operations. Secondly, the neural network weights can be either fixed or trainable. We will discuss how trainable STFT kernels improve the frequency prediction accuracy in Section~\ref{subsec:trainable}. 

\hspace{11pt} 

\noindent \textbf{nnAudio API} The STFT is implemented in nnAudio as the function \nbh{Spectrogram.STFT()}, with default arguments: n\_fft = 2048, freq\_bins = None, hop\_length = 512, window = `hann', freq\_scale = `no', center = True, pad\_mode = `reflect', fmin = 50, fmax = 6000, sr = 22050, trainable = False. This function has an optional argument \nbh{freq\_scale} which allows the user to choose either a linear or a logarithmic frequency bin scale.

\subsection{Mel spectrogram}
The Mel frequency scale was proposed by Stevens in 1937 as an attempt to quantify pitch such that equal differences in Mel-scale pitch correspond to equal differences in perceived pitch, regardless of the frequency in Hertz \cite{Stevens1937ASF}. 
In addition to the original Mel scale proposed by Stevens et al., there were several other attempts to obtain a revised version of the Mel scale~\cite{Stevens1940TheRO, fant1949analys, koening1949new}. Therefore, there is not a single ``right'' formula for Mel scale, as various different formulae coexist in the literature \cite{umesh1999fitting}.  The traditional frequency to Mel scale conversion is the one mentioned in O'Shaughnessy's book~\cite{o1987speech} which was implemented in the HTK Speech Recognition toolkit~\cite{young2002htk} as~\eqref{mel-o}), shown below,

\begin{equation}
m=2595 \text{log}_{10}\left(1+\frac{f}{700} \right)
\label{mel-o}
\end{equation}
We refer to this form as `htk' later on. Equation \eqref{mel-stanley} shows another form that is being used in the Auditory Toolbox for MATLAB~\cite{slaney1998matlab} and \nbh{librosa} (a python audio processing library)~\cite{mcfee2015Librosa}. This form is quasi-logarithmic, meaning that the frequency to Mel scale conversion is linear in the low frequency region (usually the breaking point is set to 1,000Hz), and logarithmic in the high frequency region (after the breaking point). The default Mel scale in \nbh{librosa} is in the form of \eqref{mel-stanley}, but it is possible to change it to the form defined in \eqref{mel-o} by setting the \nbh{htk} argument to \nbh{True}.

\begin{equation}
    m=
    \begin{cases}
      \frac{3f}{200}, & \text{if}\ 0 \text{Hz} \leq f\leq1000 \text{Hz} \\
      \frac{3000}{200} + \frac{27\ln{(f/1000)}}{\ln{6.4}}, & \text{if}\ f \ge 1000\text{Hz}
    \end{cases}
    \label{mel-stanley}
\end{equation}

\Figure[t](topskip=0pt, botskip=0pt, midskip=0pt)[width=85mm]{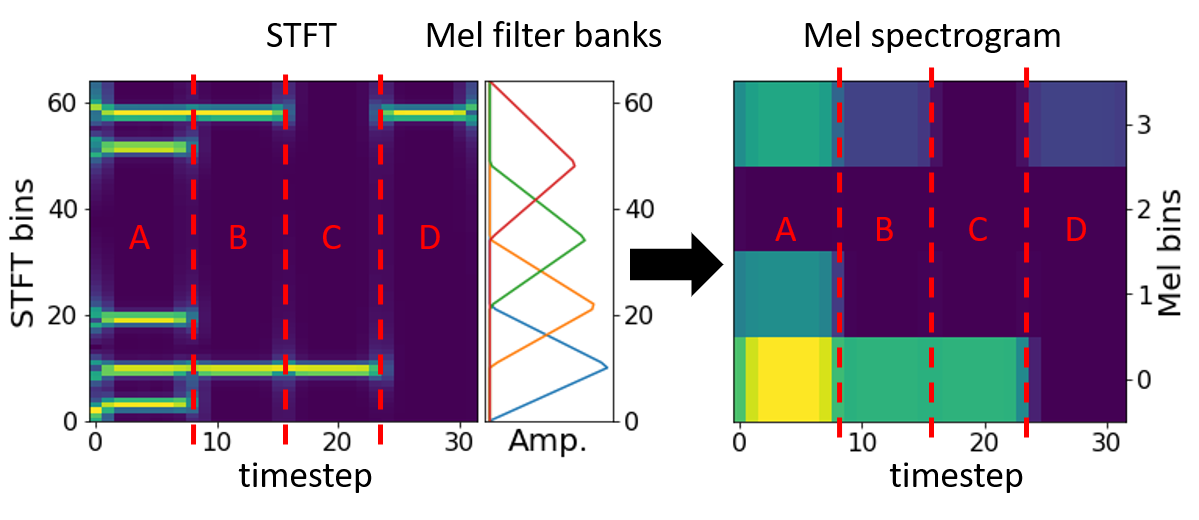}
{Mel spectrogram obtained by combining the STFT result (65 frequency bins) with 4 Mel filter banks. \label{fig: Mel spectrogram}}

Once we have the frequency to Mel scale conversion, we can create Mel filter banks (for details on the computation of Mel filter banks the reader is referred to~\cite{Davis1980ComparisonOP}) that are multiplied to each timestep of the STFT result to obtain a Mel spectrogram~\cite{rabiner2011theory}. An example of this conversion is shown in  Figure \ref{fig: Mel spectrogram}), which depicts the STFT and Mel-scale spectrograms of a signal that starts with five pure tones at 25Hz, 75Hz, 150Hz, 400Hz, and 450Hz (shown in region A). After 0.25 seconds, three of the tones stop, leaving only 2 tones at 75Hz and 450Hz (shown in region B). After another 0.25 seconds, only the 75Hz tone remains (Region C), and finally it ends with a single 450Hz tone (Region D). The STFT spectrogram is shown in the left hand side of Figure \ref{fig: Mel spectrogram}. In this example, the window size for the STFT is 128 samples, which would generate a spectrogram with 128 frequency bins. The complete spectrogram contains redundant information due to symmetry, therefore only 65 bins are used in the final STFT result. The hop size for STFT is 32 samples, which equals a quarter of the window size. To obtain a Mel spectrogram with four Mel bins, we need to have four Mel filter banks. The basis functions of a Mel filter bank are triangular in shape and the kernel that converts from the raw STFT to the Mel-spectrogram maps groups of DFT bins to a single Mel bin. 

The exact mapping for the example in Figure \ref{fig: Mel spectrogram} is shown in Table \ref{tab:mel}. There are five frequency components in region A, the three frequency components corresponding to $25$~Hz, $75$~Hz, and $150$~Hz will be mapped to Mel bin zero. Since the Mel filter banks are overlapping with each other, the frequency component $150$~Hz will also be mapped to Mel bin one. While the two high frequency components $400$~Hz and $450$Hz will only be mapped to Mel bin three. By the same logic, the Mel filter banks are multiplied with each of the STFT timesteps to obtain the Mel spectrogram.

nnAudio's implementation of Mel spectrogram calculation in \nbh{PyTorch} is relatively straightforward. We obtain the STFT results by our \nbh{PyTorch} 1D convolutional neural network discussed in Section~\ref{subsec:STFT}, and then we use Mel filter banks which were obtained from \nbh{librosa}. The values of the Mel filter banks are used to initialize the weights of a single layer fully connected neural network, and each time step of the magnitude STFT is fed forward into this fully connect layer initialized with Mel weights.  The Mel filter banks only need to be created once when initializing the neural network, These weights can be set as trainable or remain fixed,  much like the neural network implementation of STFT as discussed in Section~\ref{subsec:STFT}. Figure~\ref{fig: Mel model} shows the schematic diagram of our \nbh{PyTorch} implementation of the Mel spectrogram calculation.

\hspace{11pt} 

\noindent \textbf{nnAudio API} nnAudio implements the Mel spectrogram layer as \nbh{Spectrogram.MelSpectrogram()}, with default arguments: sr = 22050, n\_fft = 2048, n\_mels = 128, hop\_length = 512, window = `hann', center = True, pad\_mode = `reflect', htk = False, fmin = 0.0, fmax = None, norm = 1, trainable\_mel = False, trainable\_STFT = False.

\Figure[t](topskip=0pt, botskip=0pt, midskip=0pt)[width=\linewidth]{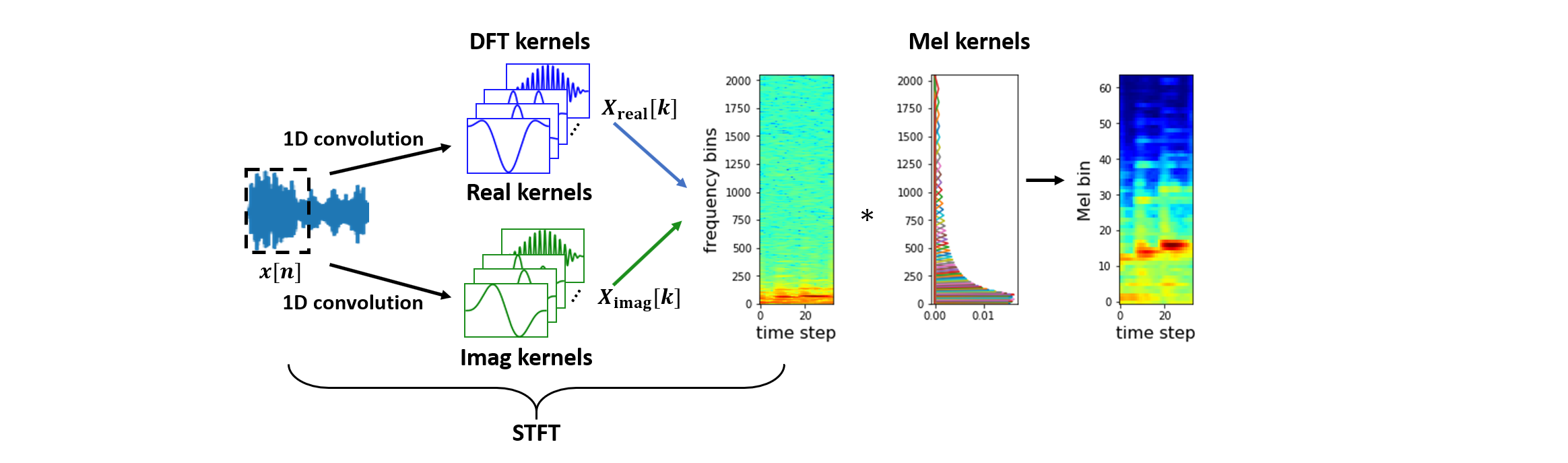}
{nnAudio's neural network based implementation for Mel spectrograms. The STFT window size is 4,096 and the number of Mel filter banks is 64 in this example.} \label{fig: Mel model}

\begin{table}\centering
\caption{Mapping from frequency bins to Mel bins for the example in Figure \ref{fig: Mel spectrogram}. The bin indexing starts with 0.}
\label{tab:mel}
\setlength{\tabcolsep}{3pt}
\begin{tabular}{lcc}
\toprule
Frequency bins& 
Corresponding & 
Mel bins\\
&frequencies& (htk version)\\
\midrule
$1-21$& $0\text{Hz} \ \text{to}\ 168.4$Hz & 0 \\
$11-34$& $79.8\text{Hz} \ \text{to}\ 267.3$Hz & 1 \\
$22-48$& $168.4\text{Hz} \ \text{to}\ 377.4$Hz & 2 \\
$35-64$& $267.3\text{Hz} \ \text{to}\ 500$Hz & 3 \\
\bottomrule
\end{tabular}
\label{tab1}
\end{table}

\subsection{Constant-Q transform}
\subsubsection{A Quick overview of the constant-Q Transform (1992 version)}
The relationships between frequencies of musical notes are logarithmic. The frequency of a musical note doubles for every one-octave increase in pitch. In order to reflect the musical relation between notes effectively on the spectrogram, it is helpful to use a logarithmic frequency scale. One naive solution is to modify the frequencies of the basis functions of the discrete Fourier transform so that the centre frequencies of the bins form a geometric series ~\cite{haines1988logarithmic}. There are, however, numerous problems with this approach.

First, it is well-known that the standard DFT basis functions of length $N$ form an orthogonal basis for the space of all complex vectors of length $N$. The orthogonality of the basis guarantees that the DFT is an energy-preserving transformation. In other words, the magnitude of the transformed output is exactly equal to the magnitude of the input. This is important because it means that we can determine the volume of the input signal simply by looking at the magnitude of the the DFT output. If we modify the frequencies of the basis vectors, they become non-orthogonal and therefore the relationship between input and output energy becomes much more complicated.

A second consequence of using unevenly spaced basis vectors in the DFT is that at the upper end of the spectrum, where the vectors are farthest apart, there will be wide gaps between frequency bins. If we insist on using a set of only $N$ vectors as the basis, these gaps are so wide that high frequency tones lying between bins will not be detected at all.
The lack of frequency resolution in the high end can be remedied by increasing the number of basis vectors beyond $N$, but doing so leads to an excessive density on the low frequency end of the spectrum. Since the width of each bin is constant with respect to frequency, this results in significant overlap between bins in the low end. In frequency ranges with significant overlap between bins, the energy shown in the transformed output is exaggerated with respect to the actual energy in the input signal.

The challenges mentioned above are the motivation for the design of the constant-Q transform, first proposed by Brown in 1991 as a modification of the discrete Fourier transform~\cite{brown1991calculation} where the window size $N_{k_{cq}}$ scales inversely proportional to the centre frequency of the CQT bin $k_{cq}$ to maintain a fixed number of cycles for sine and cosine within the window. Since the width of each bin is inversely proportional to the length of its basis vector, the width of each CQT frequency bin expands proportional to the space between bins so that there are no gaps between bins at the upper end and no excessive overlap between bins at the lower end of the spectrum.

In signal processing, the letter Q~\cite{q-factor}, which stands for Quality, indicates the centre frequency divided by bandwidth of a filter. There are many types of filters for which the term bandwidth is applied and correspondingly there are various different definitions of the bandwidth and of Q. In the context of the CQT, Q is defined to be the number of cycles of oscillation in each basis vector. The corresponding equation for Q is shown in \eqref{Q}, where $b$ is the number of bins per octave. Once $Q$ is known, we can calculate the window size $N_{k_{cq}}$ for each of the bin $k_{cq}$ by \eqref{N_k}. The equation for CQT is very similar to the DFT, with the varying index $k$ replaced by $Q$ and fixed window size $N$ replaced by varying window size $N_{k_{cq}}$ as shown in \eqref{CQT}. Despite the fact that constant-Q transform (CQT) has a similar concept with logarithmic frequency DFT, in which both of them have a logarithmic frequency scale, they are not the same. CQT maintains a constant frequency resolution by keeping a constant Q value while logarithmic frequency STFT has a varying Q. The subtle differences between the CQT and logarithmic frequency scale STFT can be seen in Figure~\ref{fig: performance_1}, ~\ref{fig: performance_2}, and,~\ref{fig: CQT_compare}

\begin{equation}
    Q=(2^{\frac{1}{b}}-1)^{-1}
    \label{Q}
\end{equation}

\begin{equation}
    N_{k_{cq}}=   \text{ceil}{\left(\frac{s}{f_{k_{cq}}}\right)Q}
    \label{N_k}
\end{equation}

\begin{equation}
    X^{cq}[k_{cq}]= \sum_{n=0}^{N_{k_{cq}}-1}x[n]\cdot e^{-2\pi i Q \frac{n}{N_{k_{cq}}}}
    \label{CQT}
\end{equation}

\subsubsection{CQT using neural networks}
The naive implementation of CQT consists of looping through all of the kernels one by one, and calculating the dot-product between the kernel $e^{-2\pi Q/N_k}$ and the input signal $x$~\cite{brown1991calculation}. This type of implementation, however, is not feasible for our 1D convolution approach. Since most of the neural network frameworks only support a fixed kernel size across different channels for 1D convolutional neural network, if we have 84 CQT kernels, we need 84 convolutional networks to include all the kernels.

\Figure[h](topskip=0pt, botskip=0pt, midskip=0pt)[width=\linewidth]{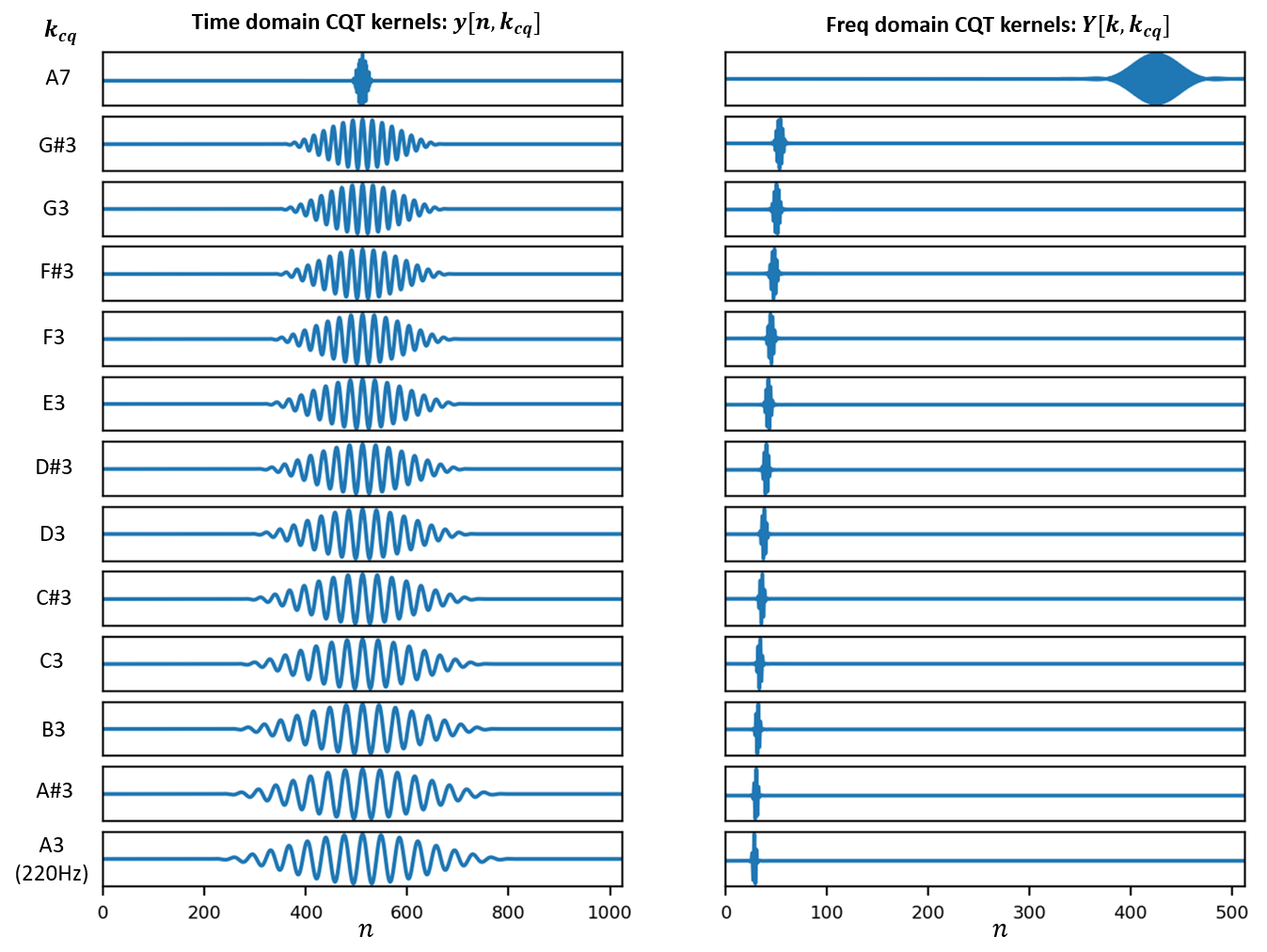}
{An example of CQT kernels whereby the number of bins per octave is set to 12. The x-axis shows the time in digital samples ($n$). Each CQT kernel has a frequency that corresponds to a musical pitch. Only the real components of $y[n, k_{cq}]$ and $Y[k, k_{cq}]$ are shown here.  \label{fig: CQT_kernels}}

Youngberg and Boll~\cite{youngberg1978constant} first proposed the concept of CQT in 1978. Brown later proposed an efficient way to calculate CQT in 1992~\cite{brown1992efficient}. The trick is to use Parseval's equation~\cite{Oppenheim1989DiscretetimeSP} as shown in \eqref{Parseval}, where $a[n]$ and $b[n]$ are arbitrary functions in the time domain, and $A[k]$ and $B[k]$ are the frequency domain versions of $a[n]$ and $b[n]$ respectively. If we define $X[k]$ and $Y[k]$ as the DFT of input $x[n]$ and kernel $e^{-2\pi Q/N_{k_{cq}}}$ respectively, then this approach converts both $x[n]$ and $e^{-2\pi iQ\frac{n}{N_{k_{cq}}}}$ to $X[k]$ and $Y[k]$ respectively in the frequency domain, and then multiplies them together to get the approximated CQT as shown in \eqref{CQT1992}. It should be noted that both $X[k]$ and $Y[k]$ are  matrices with complex numbers, and $N$ is the longest window size for the CQT kernels, which is equal to the length of the kernel with the lowest frequency. Also, $Y[k]$ is a sparse matrix in this case. Figure \ref{fig: CQT_kernels} shows an example of the CQT kernels in time domain and frequency domain respectively. The bottom and top kernels correspond to the musical note A3 ($220$Hz) and A7 ($3520$Hz) respectively, with 12 bins per octave and a sampling rate of $8000Hz$, there are 60 bins in total. Only the real components for the kernels are shown in Figure \ref{fig: CQT_kernels}, but readers should note that $y[n,k_{cq}]$ is a matrix with complex numbers, and each row of $y[n,k_{cq}]$ is transformed to a row of $Y[k,k_{cq}]$ by Fast Fourier Transform (FFT). Therefore, the frequency domain CQT kernels consist also of a matrix with complex numbers.

\begin{equation}
    \sum_{n=0}^{N_k-1}a[n]b[n] = \frac{1}{N}\sum_{k=0}^{N-1}A[k]B[k]
    \label{Parseval}
\end{equation}

\begin{equation}
    X^{cq}[k_{cq}]= \sum_{n=0}^{N_{k_{cq}}-1}x[n]\cdot e^{-2\pi i Q \frac{n}{N_{k_{cq}}}} = \frac{1}{N}\sum_{k=0}^{N-1}X[k]Y[k, k_{cq}]
    \label{CQT1992}
\end{equation}

Using Brown and et al. definition for CQT, the conversion from the time domain input $x[n]$ to $X[k]$ can be done with a 1D convolutional neural network. The DFT basis vectors will be the kernels for the neural network. Since there is a real part and an imaginary part to the DFT kernels, we need two 1D convolutional neural networks, one network for the real part kernels, and another network for the imaginary part kernels. We can perform the DFT using the same procedure as described in Section~\ref{subsec:STFT} for the STFT. Then each time step of the STFT result $X[k]$ is multiplied with the same CQT kernels $Y[k, k_{cq}]$. Therefore, the CQT kernels only need to be created once as part of the initialization for the STFT 1D convolutional nerual network. A CQT matrix $X^{cq}[k_{cq}]$, with real and imaginary part, is obtained in the end, and the final CQT ouput is calculated by taking the element-wise magnitude $\text{abs}{X^{cq}[k_{cq}]}$.

There is a major flaw for this approach unfortunately. If the number of octaves is large enough, and the CQT kernels start at a low frequency, the CQT kernels will be very huge. For example, if we want to cover 88 notes (from A0 to C8 as the range for a piano) with a sampling rate of $44100$ and 24 bins per octave, then the longest time domain CQT kernel window size is 54,727 according to \eqref{N_k}. When rounding this up to the next power of $2$, the window size will be 65,536. It has been shown that windows whose size are a power of 2 work better for FFT~\cite{rabiner1969chirp}.  Even though FFT has not been implemented in nnAudio, we will still follow these recommendations for existing CQT implementations so that we can compare them with our implementation directly. By transforming time domain CQT kernels to frequency domain kernels, we discard half of the kernel length due to symmetry. Therefore, the longest frequency domain CQT kernel has a length of 32,768. With 88 piano keys and 24 bins per octave, the CQT kernels would have a size of (176, 32768). This also implies that the window size for the STFT would be 32,768, which is extremely long, making this implementation inefficient for huge CQT kernels with a low frequency. In the following sections, we will discuss how to implement a more efficient version of CQT by using a downsampling method~\cite{brown1991calculation}.

\hspace{11pt} 

\noindent \textbf{nnAudio API} Despite its inefficiency, we still provide this function for research purposes. It can be executed in nnAudio via the function \nbh{Spectrogram.CQT1992}, with default arguments: sr = 22050, hop\_length = 512, fmin = 220, fmax = None, n\_bins = 84, bins\_per\_octave = 12, norm = 1, window = `hann', center = True, pad\_mode = `reflect', device = ``cuda:0".

\Figure[h](topskip=0pt, botskip=0pt, midskip=0pt)[width=\linewidth]{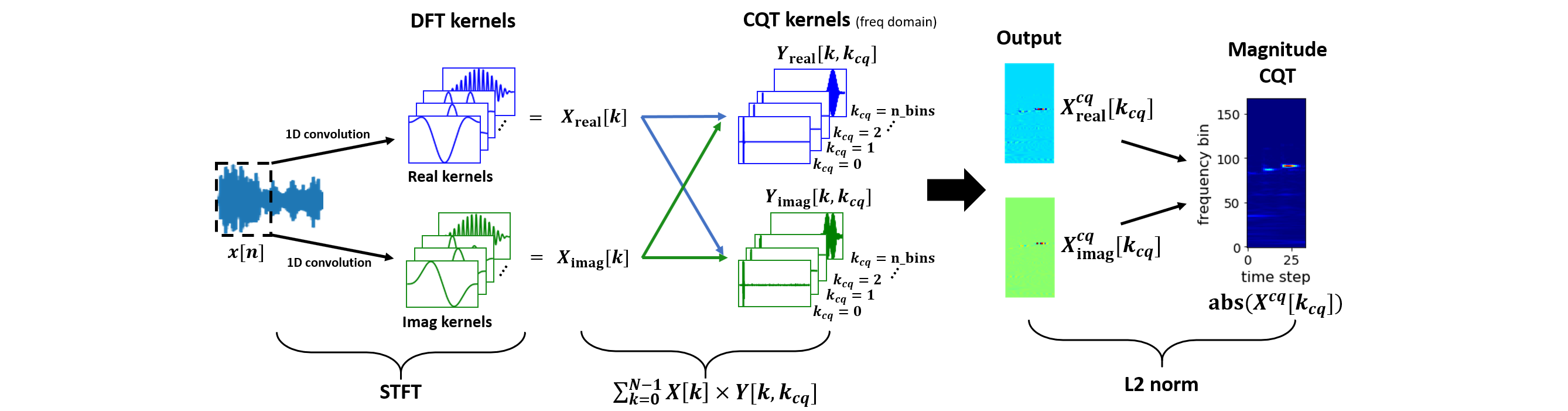}
{nnAudio's implementation of the 1992 version of CQT~\cite{brown1992efficient} by using a 1D convolutional neural network. The DFT kernels and CQT kernels only need to be initialized once and can be reused. \label{fig: CQT_torch}}

\subsubsection{Downsampling}
We will discuss how to do downsampling with a neural network before we move on to the downsampling method used in the computation of the CQT. In order to downsample the input audio clips by a factor of two without aliasing, a low pass filter is required so that any frequencies above the downsampled Nyquist frequency will be filtered out first, before performing the actual downsampling. This is performed by a technique called Finite impulse response filtering (FIR). FIR refers to the convolution of an input signal with a filter kernel using the same formula shown earlier in \eqref{convolutionDefinition}. This type of filtering can be implemented efficiently using a convolutional neural network.
The definition of FIR is shown in~\eqref{FIR}, where $x[n-i]$ is the input signal at time step $n$, $b\_i$ is the FIR filter. 

To downsample, we first design the low-pass FIR filter kernel using the Window Method~\cite{rajput2012implementation}, which is implemented in \nbh{SciPy} as the function \nbh{scipy.signal.firwin}. To achieve a steep cutoff at the Nyquist frequency we set the passband of the filter to end at 0.4995 and the stopband to start at 0.5005 times the Nyquist frequency. These values were chosen so as to achieve a steep cutoff. The impulse response and frequency response of our antialiasing filter is shown in Figure~\ref{fig:low-pass-filter}. This filter is used as the kernel of the downsampling component of our convolutional neural network. An effective antialiasing filter design is important for the implementation of the CQT, which we explain in the following section.

\Figure(topskip=0pt, botskip=0pt, midskip=0pt)[width=80mm]{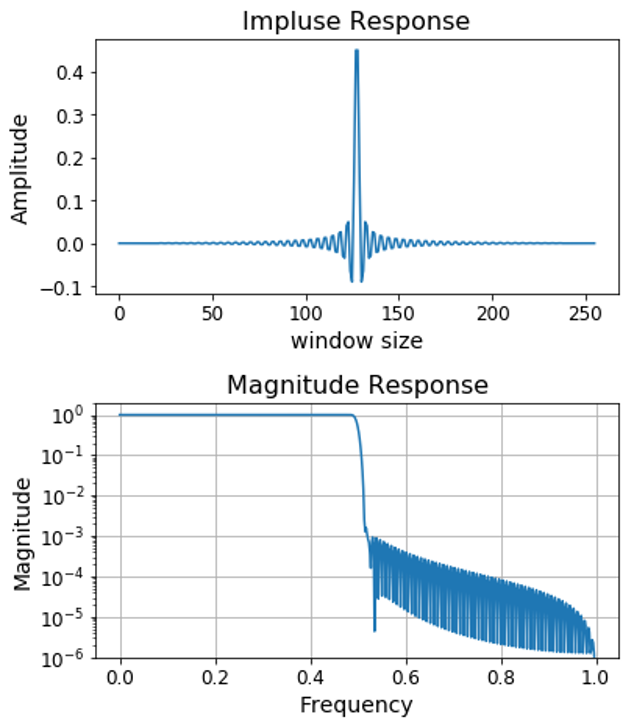}
{Impulse response and magnitude frequency response for the antialiasing filter. This filter forms the kernel for the 1D convolutional neural network that performs downsampling in nnAudio's CQT computation. \label{fig:low-pass-filter}}

\begin{equation}
    y[n] = \sum_{i=0}^{N}b_i\cdot x[n-i]
    \label{FIR}
\end{equation}

\subsubsection{Constant-Q transform (2010 version)}

The constant-Q transform uses basis vectors of varying lengths. The basis kernels for the lowest frequencies are several orders of magnitude longer than the high frequency kernels. Since low frequency audio signals can be accurately represented with lower sample rates, we can compute the lower frequency components of the CQT more efficiently by downsampling the input and using correspondingly shorter filter kernels. This technique is described in detail in~\cite{schorkhuber2010constant, brown1991calculation}. Only one octave of CQT kernels are created under this approach, and these CQT kernels usually start from the highest octave due to the short window size as described in \eqref{N_k}. By doing so, the computational complexity can be reduced. When applying the CQT kernels (of this highest octave) to the frequency domain input $X[k]$, only the CQT result for the highest octave is obtained. After that, we downsample the input by a factor of two, and apply the same CQT kernels to this new input to obtain the CQT result for the next octave down. The same process is repeated until the desired number of octaves is processed. In this approach, the CQT kernels are kept the same while the input audio is being downsampled recursively. By referring to \eqref{N_k}, $N_{k_{cq}}$ and $Q$ are constant. When we downsample the audio by a factor of 2, $s$ is reduced by half. In order to keep $N_{k_{cq}}$ and $Q$ constant, $f_{k_{cq}}$ must be reduced by half. Physically, it means the CQT output obtained by same CQT kernels relative to a downsampled audio with factor $2^\alpha$ is $\alpha$ octave lower than the original audio, where $\alpha \in [1,2,3,...]$ is a positive integer. Figure~\ref{fig: CQT2010} shows the schematic diagram for this implementation. Each downsampled input $x_\alpha[n]$ produces the CQT result for one octave. The complete CQT result can then be obtained by appending the results for each of the octaves together.

\hspace{11pt} 

\noindent \textbf{nnAudio API} This algorithm can be executed in nnAudio via the function \nbh{Spectrogram.CQT2010}, with default arguments: sr = 22050, hop\_length = 512, fmin = 32.70, fmax = None, n\_bins = 84, bins\_per\_octave = 12, norm = True, basis\_norm = 1, window = `hann', pad\_mode = `reflect', earlydownsample = True, device = `cuda:0'.

\Figure(topskip=0pt, botskip=0pt, midskip=0pt)[width=\linewidth]{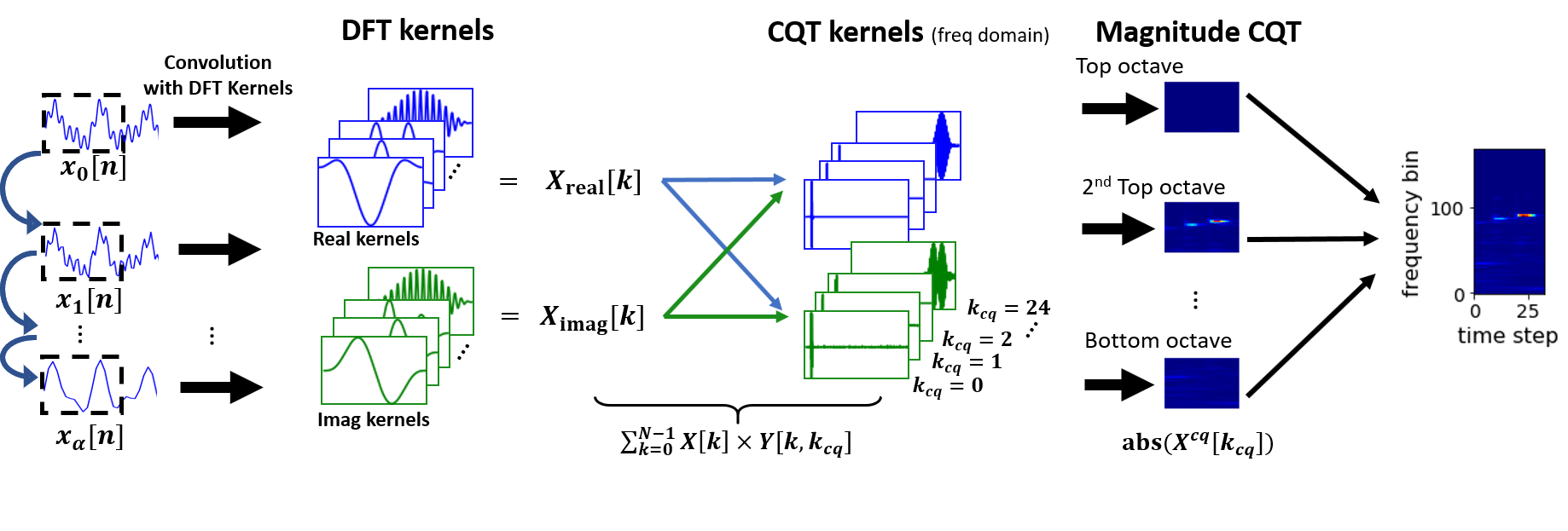}
{Schematic diagram of the 2010 version of CQT~\cite{schorkhuber2010constant, brown1991calculation} using the recursive downsampling method. The kernels  only need to be initialized once and can be reused over and over again. \label{fig: CQT2010}}

\subsubsection{CQT with time domain kernels}
\label{subsec:CQTv2}
When \citet{brown1992efficient} proposed their more efficient algorithm in 1992, they were facing limitations in computer memory. The time domain CQT kernels form a very large, dense matrix. Storing a matrix like this requires a lot of memory. When converting time domain CQT kernels into frequency domain kernels, the dense matrix becomes a sparse matrix. Storing this sparse matrix using either the compressed sparse row (CSR) format or the compressed sparse column (CSC) algorithm is more memory efficient than storing a dense matrix. Therefore, by converting the time domain CQT kernels to the frequency domain, the same information is retained, while requiring less memory to store it.

With modern technology, memory is no longer an issue. Thus, it is no longer necessary to convert the time domain CQT kernels to frequency domain kernels. By doing so, we remove a computationally heavy step, thus improving the CQT computation speed. Both the 1992 version of CQT and the 2010 version of CQT can benefit from this modification. The resulting modified implementation is shown in Figure~\ref{fig: CQT2010v1} and \ref{fig: CQT2010v2}. The improvement in computational speed is reported as CQT1992v2 and CQT2010v2 respectively for each algorithm in Figure~\ref{fig:speed}.

\hspace{11pt} 

\noindent \textbf{nnAudio API} The improved version of CQT1992 is implemented as \nbh{Spectrogram.CQT1992v2()} with the default parameters: sr = 22050, hop\_length = 512, fmin = 32.70, fmax = None, n\_bins = 84, bins\_per\_octave = 12, norm = 1, window = `hann', center = True, pad\_mode = `reflect', trainable = False, output\_format = `Magnitude', device = `cuda:0'. 

The improved version of CQT2010 can be executed in nnAudio via the function \nbh{Spectrogram.CQT2010v2()} with default parameters: sr = 22050, hop\_length = 512, fmin = 32.70, fmax = None, n\_bins = 84, bins\_per\_octave = 12, norm = True, basis\_norm = 1, window = 'hann', pad\_mode = 'reflect', earlydownsample = True, device = 'cuda:0'.

\Figure(topskip=0pt, botskip=0pt, midskip=0pt)[width=80mm]{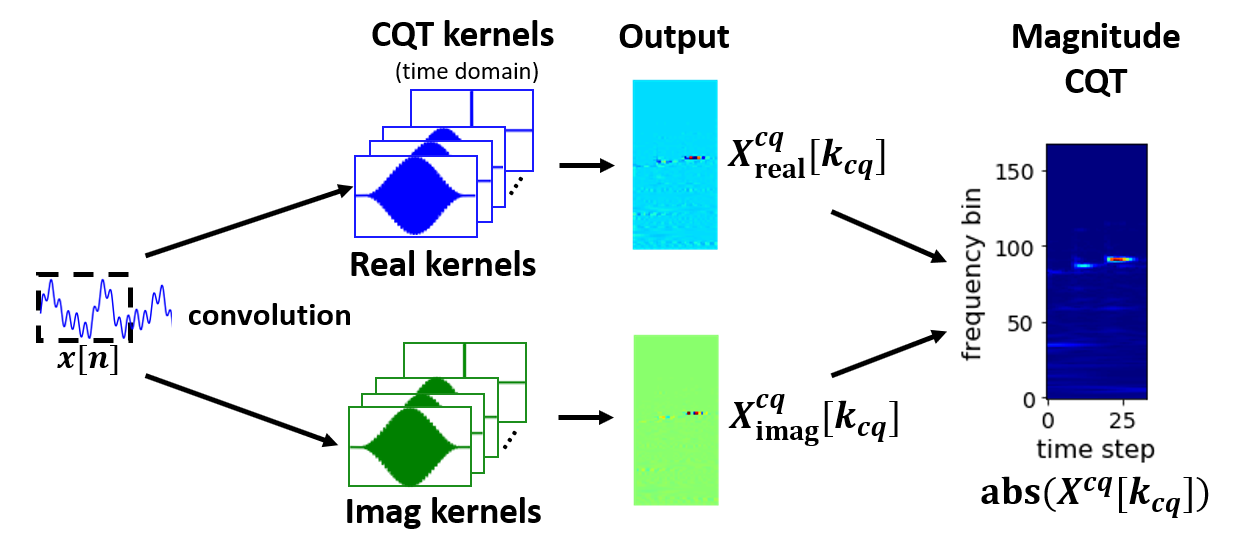}
{A schematic diagram showing our proposed improvement of the CQT1992 algorithm that uses time domain CQT kernels instead of frequency domain kernels, which require less computational steps compared to the original algorithm as shown in Figure~\ref{fig: CQT_torch}\label{fig: CQT2010v1}}

\Figure(topskip=0pt, botskip=0pt, midskip=0pt)[width=80mm]{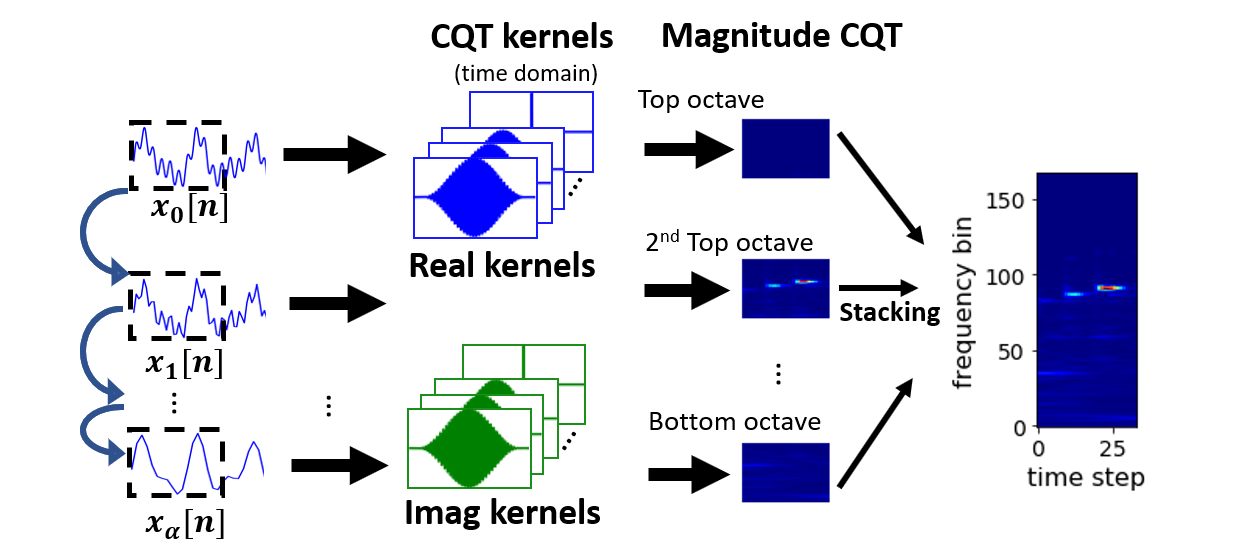}
{A schematic diagram showing our proposed improvement of the CQT2010 algorithm that uses only time domain CQT kernels. Note that the output of the convolution between the audio input and the CQT kernels is still a complex number, even though this is not shown in the figure for simplicity. \label{fig: CQT2010v2}}

\section{Experimental results}\label{sec: Result}
In this section, we analyse the speed and the accuracy of the proposed time domain to frequency domain transformations. We compare our \nbh{PyTorch} implementation, nnAudio, with the existing audio processing library \nbh{librosa}~\cite{mcfee2015Librosa}. More specifically, our STFT implementation is compared to \nbh{librosa.stft}, Mel Spectrogram to \nbh{librosa.feature.melspectrogram}, and CQT to \nbh{librosa.cqt}. In the first subsection, we compare the speed required to  process 1,770 audio files in wav format. In the second subsection, we focus on testing the correctness of the resulting spectrograms.  In what follows, the different implementations for CQT,  namely CQT1992v2 and CQT2010v2, will be discussed individually. These are the implementations that directly use time domain CQT kernels as mentioned in Section~\ref{subsec:CQTv2}. For the sake of easy reference, Mel spectrogram will be referred as MelSpec below.

\subsection{Speed}\label{subsec: speed}

\subsubsection{Setup}
We use the MAPS dataset~\cite{emiya2010maps} to benchmark nnAudio. A total of 1,770 wav files from the \nbh{AkPnBcht/UCHO/} folder were used for the benchmark. We discard the first 20,000 samples (which is equivalent to 0.45 seconds under the 44.1kHz sampling rate) from each audio excerpt in order to remove the silence. Each of the audio excerpts are kept the same length (80,000 samples) by removing the excessive samples in the end. Their final length is equivalent to 1.8 seconds. The audio excerpts are stored as an array with shape $1,770 \times 80,000$. The goal of the speed test is to convert this array of waveforms into an array of spectrograms while maintaining the order of the audio excerpts. We conducted this test in a DGX station with CPU: Intel(R) Xeon(R) CPU E5-2698 v4 @ 2.20GHz and 4 Tesla v100 32Gb GPUs. During the test, we compare the speed of our proposed nnAudio toolkit to one of the popular signal processing libraries, \nbh{librosa}~\cite{mcfee2015Librosa}. Although \nbh{Essentia} is reported to be faster than \nbh{librosa} in terms of audio processing speed~\cite{moffat2015evaluation}, our experimental results show that \nbh{Essentia} is slower than \nbh{librosa}. (It takes \nbh{Essentia} 30 seconds to finish the STFT task and 180 seconds to finish the CQT task.) One possible reason is that \nbh{Essentia} only supports the versions of STFT and CQT without moving window. Therefore it can only produce spectrums, not spectrograms. To obtain the spectrograms, we first need to cut the input audio into small audio segments and then apply the CQT or STFT on each of these segments. This is done using extra nested for loops, which could cause a slower speed in \nbh{Essentia}. On top of that, \nbh{Essentia} does not support MelSpec, making a side by side comparison to nnAudio impossible. We therefore report on the results for \nbh{nnAudio} and \nbh{librosa} in Figure~\ref{fig:speed}.

Since our task is to transform an array of waveforms to an array of spectrogram (in the same order), \nbh{librosa} with multi-process will not work well for us. The time it takes to finish this task while maintaining the same order for the output array as the original array is longer than a plain \textbf{for} loop when multiprocessing is used. Therefore, the speed test for \nbh{librosa} is performed by using a for loop. Furthermore The performance for \nbh{librosa} can be optimized by using caching, but this option is disabled by default. To emulate the situation that most people use \nbh{librosa}, we run speed test on it with caching disabled. Even when caching is used, it only reduces the computation time for CQT by around 10 seconds. As for nnAudio, despite the fact that multiple GPUs are available on the DGX, only one GPU is used to convert the array of waveforms to the array of spectrograms. Since \nbh{PyTorch} can also be run on a CPU, we will also test this configuration of nnAudio.

\Figure(topskip=0pt, botskip=0pt, midskip=0pt)[width=80mm]{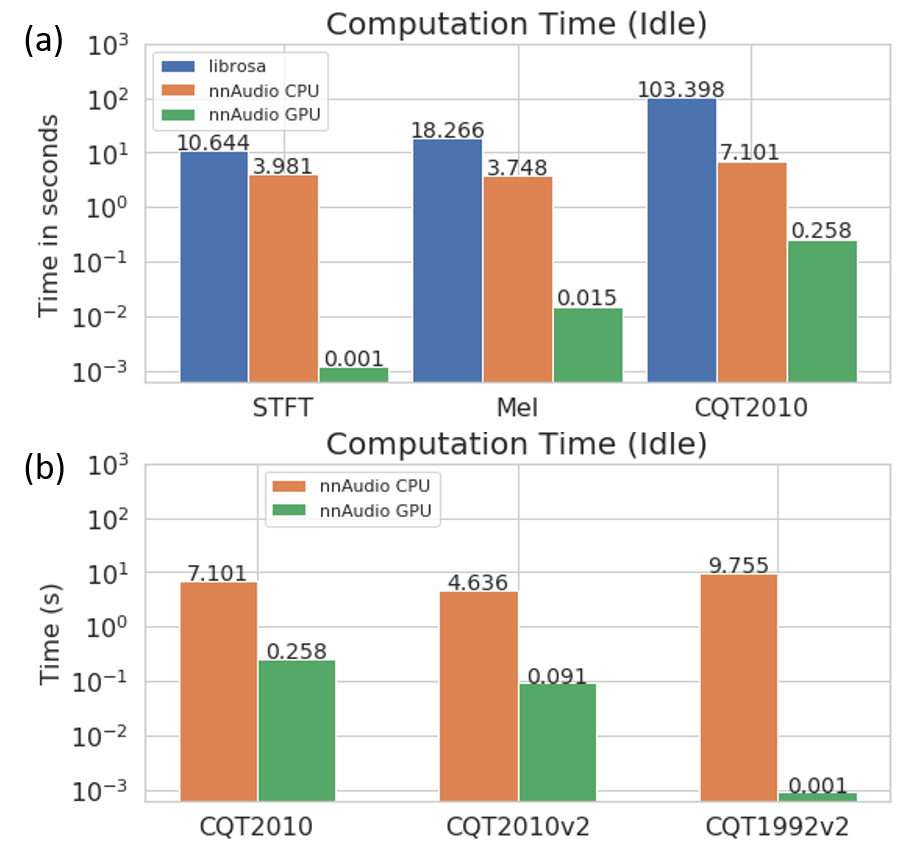}
{Processing times to compute different types of spectrograms with nnAudio GPU, nnAudio CPU, and \nbh{librosa}. \label{fig:speed}}

\begin{table}[h!]
\caption{The GPU initialization time needed for the two kernels (STFT and model specific kernel) in each nnAudio neural network model, together with the required memory. }
\centering
\label{tab: mem}
\setlength{\tabcolsep}{5pt}
\begin{tabular}{p{60pt} p{45pt} p{45pt} p{35pt}}
\toprule
Model& 
DFT kernels (in s)& 
Model kernels (in s) & Memory (in MiB)\\\midrule
STFT:   & $5.5\pm 0.1$   & N.A. & $1135$\\
n\_fft=4096 & & & \\\hdashline
MelSpec:  n\_fft=4096, n\_mels=512 & $5.5\pm 0.04$  & $0.02\pm 0.004$ & $1155$\\\hdashline
CQT1992: bins\_p\_oct=24, bins=176  & $194.9\pm 1.0$ & $5.6 \pm 0.08$ & $17505$\\\hdashline
CQT1992v2: bins\_p\_oct=24, bins=176& N.A.           & $4.9 \pm 0.08$ & $1157$\\\hdashline
CQT2010: bins\_p\_oct=24, bins=176  & $0.05\pm 0.03$ & $4.8\pm 0.08$ & $1177$\\\hdashline
CQT2010v2: bins\_p\_oct=24, bins=176& N.A            & $4.6\pm 0.05$ & $1089$\\
\bottomrule
\end{tabular}
\label{tab: GPU_util}
\end{table}

\subsubsection{Results}
Figure~\ref{fig:speed} shows the time taken to convert an array of 1,770 waveforms to an array of 1,770 spectrograms using Mel frequency scale, STFT, and CQT. It is clear from the figure that our newly proposed method is at least 100 times faster than \nbh{librosa}. We should note that, when using \nbh{PyTorch} with GPU, extra time is required to transfer the kernels from RAM to GPU memory, which only takes a few seconds. This process can be considered as part of the model initialization, the time required to initialize the models is not included in Figure~\ref{fig:speed} Table~\ref{tab: GPU_util} shows the time taken to initialize each neural network model with nnAudio. This time is influenced by the kernel sizes of STFT, MelSpec, and CQT. For STFT, a longer window size (n\_fft) results in larger STFT kernels. The same goes for MelSpec and CQT. More time is required to transfer larger kernels to GPU memory. In our experiment, an STFT window size of 4,096 is used for both STFT and MelSpec. For MelSpec, a total of 512 Mel filter banks are used. For the different implementations of CQT, the kernels start at $32.7$Hz which corresponds to the note C1, and 24 bins per octaves, covering 176 bins in total. The neural network models used by nnAudio to calculate MelSpec and CQT require operations with multiple kernels (an initial DFT kernel followed by a model specific kernel), therefore, we break the initialization time down into two steps (columns 2 and 3 in Table~\ref{tab: GPU_util}). Model kernels refer to the convolution kernels specific to each spectrogram type. For MelSpec, the model kernels are would be the Mel filter banks. For CQT, they consist of both DFT and CQT kernels. The initialization the kernels of the network only needs to be performed once. As we can observe from Table~\ref{tab: GPU_util}, CQT2010 has a much faster initialization time compared to CQT1992 (5 seconds compared to over 200 seconds). This can be explained as the bottleneck for CQT1992 in the STFT stage. If the starting frequency is too low, the CQT kernels become very long, which in turn causes a huge window size (n\_fft) for STFT. In the CQT setting used for the kernel initialization speed test (sampling rate=$44,100$Hz, minimum frequency= $32.7$Hz, bins per octaves=24, and bins=176), the longest CQT kernel is 46,020, which results in a n\_fft of 65,536 (rounding up the to nearest powers of two, $2^{16}$). To mitigate this problem, a Fast Fourier Transform (FFT) could be used instead of DFT, which will be explored in future research. Another way to prevent this problem would be to use the implementation mentioned in~\ref{subsec:CQTv2}. Once everything is loaded into the GPU memory, the computation will occur at the speed as shown in Figure~\ref{fig:speed}~(a) and Figure~\ref{fig:speed}~(b). Even when only a CPU is used, nnAudio still outperforms \nbh{librosa} and \nbh{Essentia} significantly.

As mentioned in Section~\ref{subsec:CQTv2}, converting the time domain CQT kernels to frequency domain CQT kernels is not necessary for if there is enough computing memory. In the experiment, we compare the improvement in computation speed when using the time domain CQT kernels directly. Figure~\ref{fig:speed}~(c) shows how the improved constant-Q Transform (CQT1992v2) and the improved constant-Q Transform with downsampling (CQT2010v2) further improve the computation speed. CQT2010v2 is faster than CQT2010 regardless if the CPU or GPU is used. While CQT1992v2 is extremely fast when GPU is used, the CPU version is slower than the CQT2010. Therefore, it is tempting to conclude that CQT2010v2 should be used in a computer without GPU, and CQT1992v2 should be used when GPU is available. However, there are subtle differences between the 1992 and 2010 implementation, and. under normal circumstances, CQT1992v2 is the best option among all the implementations. The subtle differences between various CQT implementations will be discussed in detail in the following subsection. Nevertherless, to ensure flexibility, \nbh{nnAudio} provides all implementations discussed above (CQT1992, CQT2010, CQT1992v2, CQT2010v2).

\subsection{Conversion output}

\subsubsection{Setup}
We use \nbh{librosa} as our benchmark to check the correctness of our implementation. The spectrograms produced by our implementation are compared to the \nbh{librosa} output by using the numpy function \nbh{np.allclose}. Four input signals, a linear sine sweep, a logarithmic sine sweep, an impulse tone, and a chromatic scaled played on a piano, are used in this study to determine the model output correctness. The chromatic piano scale is recorded with a piano instrument provided by Garritan Personal Orchestra 5\footnote{\url{https://www.google.com/search?client=firefox-b-d\&q=Garritan+Personal+Orchestra+5}} and saved as a wav file. Since adapting the time domain CQT kernels does not change the output spectrogram, the result for CQT1992 is same as CQT1992v2, and better than the result for CQT2010 and CQT2010v2. Therefore we will only report the result for the faster and better quality implementation (CQT1992v2) here.

\subsubsection{Results}
The results of the accuracy test are shown in Figures~\ref{fig: performance_1} and \ref{fig: performance_2}. The output magnitudes are displayed in a logarithmic scale so that the subtle differences can be observed easily. When looking at the results, we notice that the STFT results from \nbh{librosa} and nnAudio are very similar to each other with an absolute tolerance of $10^{-2}$ and a relative tolerance of $10^{-2}$. The same can be said for MelSpec, for which the results of both libraries are very similar, with an absolute tolerance of $10^{-3}$ and a relative tolerance of $10^{-4}$. For CQT, the absolute tolerance is $0.8$ and the relative tolerance is $2$. The CQT output for nnAudio is smoother because we are using the CQT1992v2 approach in our implementation for which downsampling is not required. Figure~\ref{fig: CQT_compare} shows the comparison between our proposed CQT1992v2, our CQT2010v2 and \nbh{librosa}'s implementation of CQT (\nbh{librosa} is using the 2010 downsampling algorithm). In the implementation of both librosa and CQT2010v2, the aliasing in the figure is due to downsampling. Although the magnitude of the aliasing is negligible, it is still observable when we use a logarithmic magnitude scale. Further study is required to determine the effects of the aliasing due to downsampling in the neural network models. The CQT1992v2 model, however, is the fastest of all proposed GPU-based CQT implementations (see Figure~\ref{fig:speed}(c)), and its output is the best among the different implementations. Therefore CQT1992v2 should be used and hence it is set as the default CQT computation algorithm for \nbh{nnAudio}.

\Figure(topskip=0pt, botskip=0pt, midskip=0pt)[width=170mm]{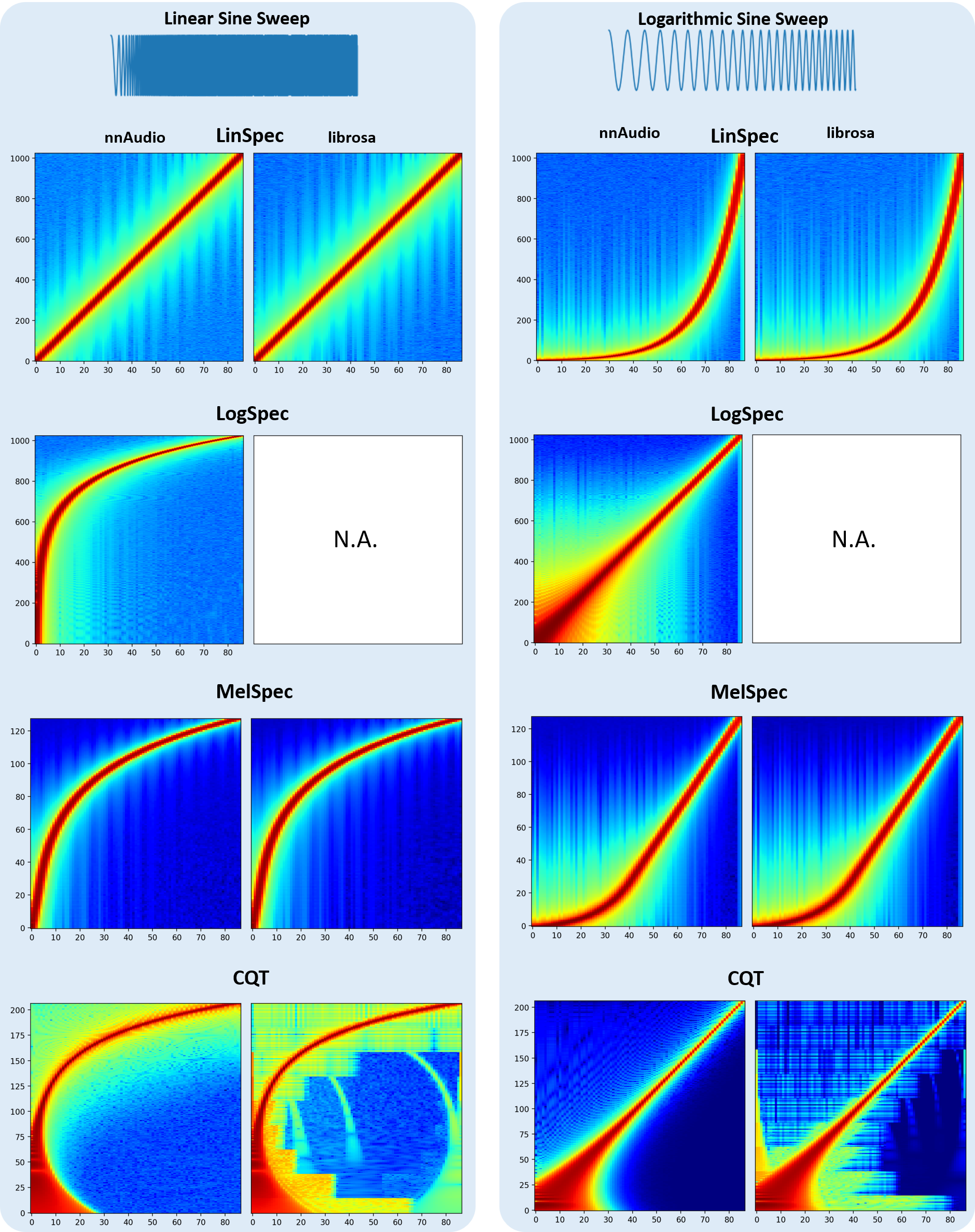}
{Comparing the outputs of \nbh{nnAudio} and \nbh{librosa} when converting a linear and logarithmic sine sweep as the inputs, \label{fig: performance_1}}
\Figure(topskip=0pt, botskip=0pt, midskip=0pt)[width=170mm]{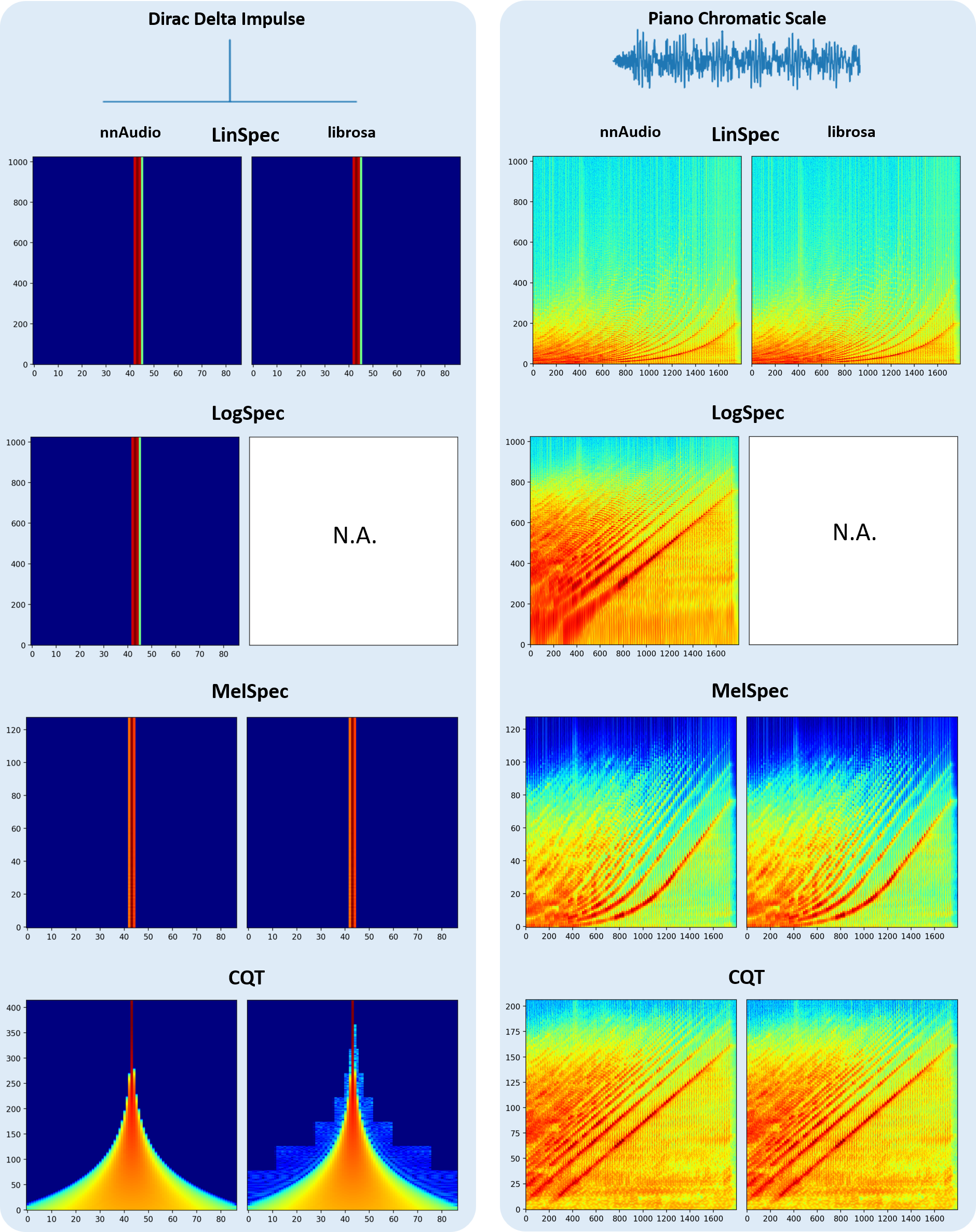}
{Comparing the outputs of \nbh{nnAudio} and \nbh{librosa} when converting an impulse tone and a chromatic piano scale. \label{fig: performance_2}}

\Figure(topskip=0pt, botskip=0pt, midskip=0pt)[width=85mm]{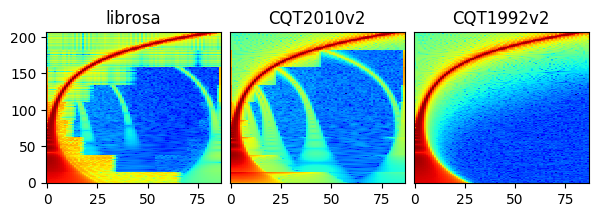}
{A visualisation of the subtle differences between \nbh{librosa}, CQT2010v2 and CQT1992v2 implementations using a logarithmic scale. CQT1992v2 yields the best result. A linear sine sweep is used as the input signal.\label{fig: CQT_compare}}

\section{Example applications}
In this section, two potential applications of nnAudio will be discussed. First, we will elaborate on using nnAudio to explore different spectrograms as the input for a music transcription model, and discuss how this process can benefit from on-the-fly GPU processing. Second, we will demonstrate that nnAudio allows the STFT kernels to be trained/finetuned, so that a better spectrogram can be obtained.

\subsection{Exploring different input representations}
In this section, we discuss one possible application, namely music transcription~\cite{benetos2013automatic, holzapfel2019automatic}. We will show that with nnAudio, one can quickly explore different spectrograms as the input for a neural network and easily choose the spectrogram that yields the best transcription accuracy.

Consider the following scenario: we want to do polyphonic music transcription, and we have picked a specific model (fully connected neural network) to tackle this task, but we want to know which input spectrogram representation would yield the best results for this task. nnAudio will enable us to easily tackle this scenario. Four types of spectrograms are explored: linear frequency scale spectrogram (LinSpec), logarithmic frequency scale spectrogram (LogSpec), Mel spectrogram (MelSpec), and CQT. In addition, each of these representations will have different parameters settings that we need to consider. For LinSpec and LogSpec, we want to explore five different sizes of Fourier kernels. For MelSpec, we want to explore four different sizes of Fourier kernels, and for each Fourier kernels, we want to further explore the number of Mel filter banks. Finally for CQT, we want to explore ten different bins per octaves. This means that there will be in total 34 different input representations. The traditional way to do this is to convert the audio clips into 34 different sets of spectrograms, and load each set to the model in order to find the best input representation, which will be a slow process that requires lots of hard disk space. We use audio recordings and their annotations from MusicNet~\cite{thickstun2017learning, thickstun2018invariances} as the waveform input in our experiment, and obtain different spectrograms with different parameter settings on-the-fly. Figure~\ref{fig: accuracy_plot} shows the transcription accuracy obtained with different input representations. The accuracy is measured by using \nbh{mir\_eval.multipitch.metrics()}. Since all these parameters affect the output shape of the spectrograms (number of bins), the result can be plotted as transcription accuracy versus number of frequency bins. It is clear from the image that the input representation and its settings has a big influence on model performance. Using nnAudio, which will enable fast comparison of different representations, will ultimately result in easier to configure and more efficient models.

With nnAudio, we were able to easily add the spectrogram calculation as the first layer of our model. This first layer is responsible for waveform to spectrogram conversion during the feedforward process. In this way we only need to store the audio clips in the original waveform, without saving extra copies of dataset for the spectrograms. In addition, nnAudio is also useful when the dataset is so large that it takes tens of hours to convert the data from waveforms to spectrograms. Once the waveforms are ready, it can be loaded batch by batch (when using \nbh{PyTorch}) and fed-forward to nnAudio, which then converts batches of waveforms into spectrograms on-the-fly. This save the user the trouble processing the original waveforms and saving them as 34 different sets of spectrogram on the hardisk. Yet, it still allows us to perform the same analysis on the results~\ref{fig: accuracy_plot}. The full details of this experiment are outside of the scope of this paper and may be published in future work. 

\Figure(topskip=0pt, botskip=0pt, midskip=0pt)[width=70mm]{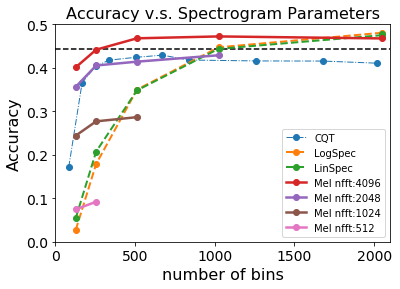}
{Performance of the four different input representations, with different parameters settings, when performing audio transcription of the audio files in the MusicNet dataset~\cite{thickstun2017learning, thickstun2018invariances}.  \label{fig: accuracy_plot}}

\subsection{Trainable Transformation Kernels}\label{subsec:trainable}
Since we implement STFT and MelSpec with a 1D convolution neural network whereby the neuron weights correspond to the Fourier kernels and Mel filter banks, it is possible to further finetune these kernels and filter banks together with the model via gradient descent. This technique is available for all transformations implemented with a neural network, but we will only focus on discussing the STFT and MelSpec in this subsection as an example.

Consider the following task: given a pure sine wave, we need to train a model that is able to return the frequency of the signal. To make this task non-trival, the STFT window size is deliberately set to a small number $64$, so that the output spectrograms have a very poor frequency resolution. The frequencies for pure sine waves are integers ranging from $200$Hz to $22,050$Hz (the Nyquist frequency). In other words, we have only $33$ frequency bins with which to resolve the entire audible spectrum from 20 Hz to 20 KHz. To conduct our experiment, we generated 10,925 pure sine waves with different frequencies (between $200$ and $22,050$Hz). For each frequency, we generate 10 different pure sine waves with different phases. In total, 109,250 pure sine waves are generated to form our dataset. 80\% of these sine waves are used as the training set, and the remaining 20\% are used as test set. We explore here if trainable kernels could improve the model accuracy. We focus on two models for predicting the frequency of the input sign wave: a fully connected network and a 2D convolutional neural network (CNN). For the fully connected network, we use one layer with one neuron and sigmoid activation. The spectrogram is flattened to a 1D vector, and used as the input to model. For CNN, two 2D convolution layers are used, with a kernel size ($4\times4$) for each layer. The final feature maps of the CNN are flattened and fed forward to a fully connected network with one neuron and sigmoid activation. nnAudio is used as the first layer of these models, to convert waveforms to either standard spectrograms, Mel spectrograms, or CQT spectrograms. We set this first layer to be trainable and compare the resulting loss to the same model with this layer is set non-trainable. As can be seen in Figure~\ref{fig: trainable_loss}, a trainable transformation layer results in a lower mean square error (MSE) for STFT, MelSpec, and CQT layers and for both Linear and CNN models.

\Figure(topskip=0pt, botskip=0pt, midskip=0pt)[width=85mm]{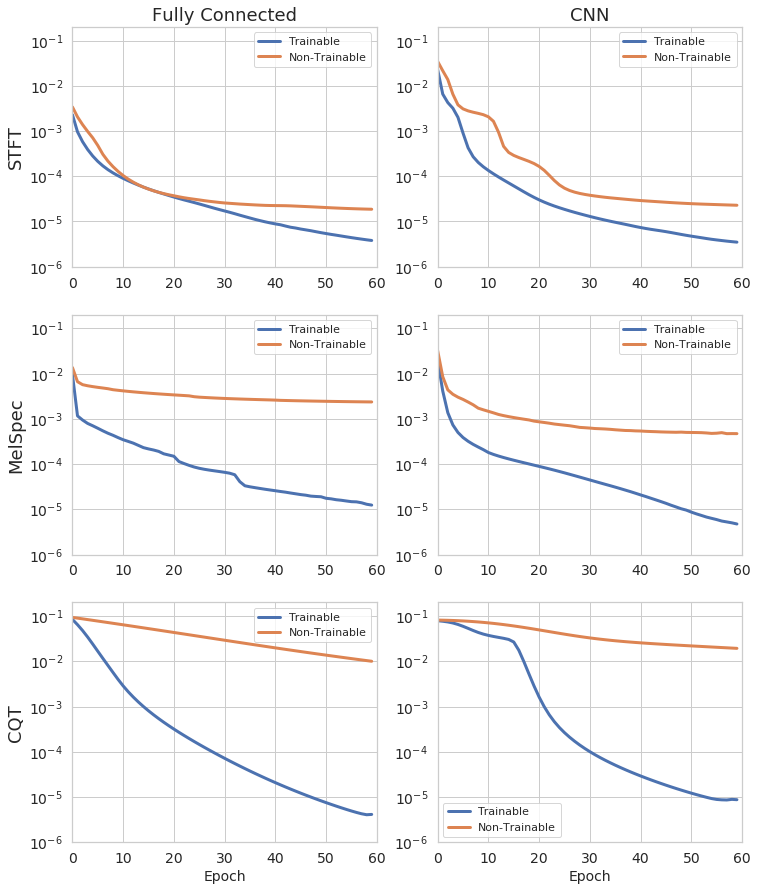}
{The evolution of loss during training for trainable and non-trainable kernels on a frequency prediction task. The models with trainable kernels consistently outperform the models with a fixed kernels. The models were trained on 87,400 pure sine waves and evaluated on 21,850 pure sine waves.\label{fig: trainable_loss}}

In order to explain how a trainable STFT, MelSpec, and CQT layer improves the prediction accuracy, we need to study the trained Fourier kernels and Mel filter banks. The first two rows in Figure~\ref{fig: Trained_STFT} show the Fourier Basis when the filter bank are $k=1,2$. Since the results for the fully connected model are quite similar to the CNN model, we will only report the results for the CNN model here. The column on the left visualizes the original Fourier kernels, and the column on the right visualizes the trained Fourier kernels. Although the overall shape of the trained Fourier kernels is similar to the original Fourier kernels, it contains some higher frequencies on top of the fundamental frequency for the kernels. These extra frequencies may allow more information to be extracted via STFT. The trained STFT spectrogram is shown in the last row of the same figure. It is clear from this figure that it has more overtone-like signals around the fundamental frequency, while the original STFT shows a very clean response for the pure sine wave input. The spectrogram obtained via the trained STFT may be able to provide clues to the neural network about the input frequency of the input signal. The same is true for the trained Mel filter banks and CQT kernels as shown in Figure~\ref{fig: Trained_Mel} and~\ref{fig: Trained_CQT}. By allowing the neural network to further train or finetune the Mel filter banks and CQT kernels, we allow a richer spectrogram to be obtained. This provides the frequency prediction models, regardless of the network architecture, with more information so as to reach a lower MSE loss.

This subsection shows that further training or finetuning the spectrogram transformation layer with nnAudio results in a lower MSE loss. Despite the fact that this analysis uses a simple, artificially generated dataset, it still provides a good example of how a higher performing end-to-end model can be obtained with a trainable transformation layer. The detailed experimental results are available on the \nbh{nnAudio} github repository\footnotemark[\ref{nnAudio}].

\Figure(topskip=0pt, botskip=0pt, midskip=0pt)[width=85mm]{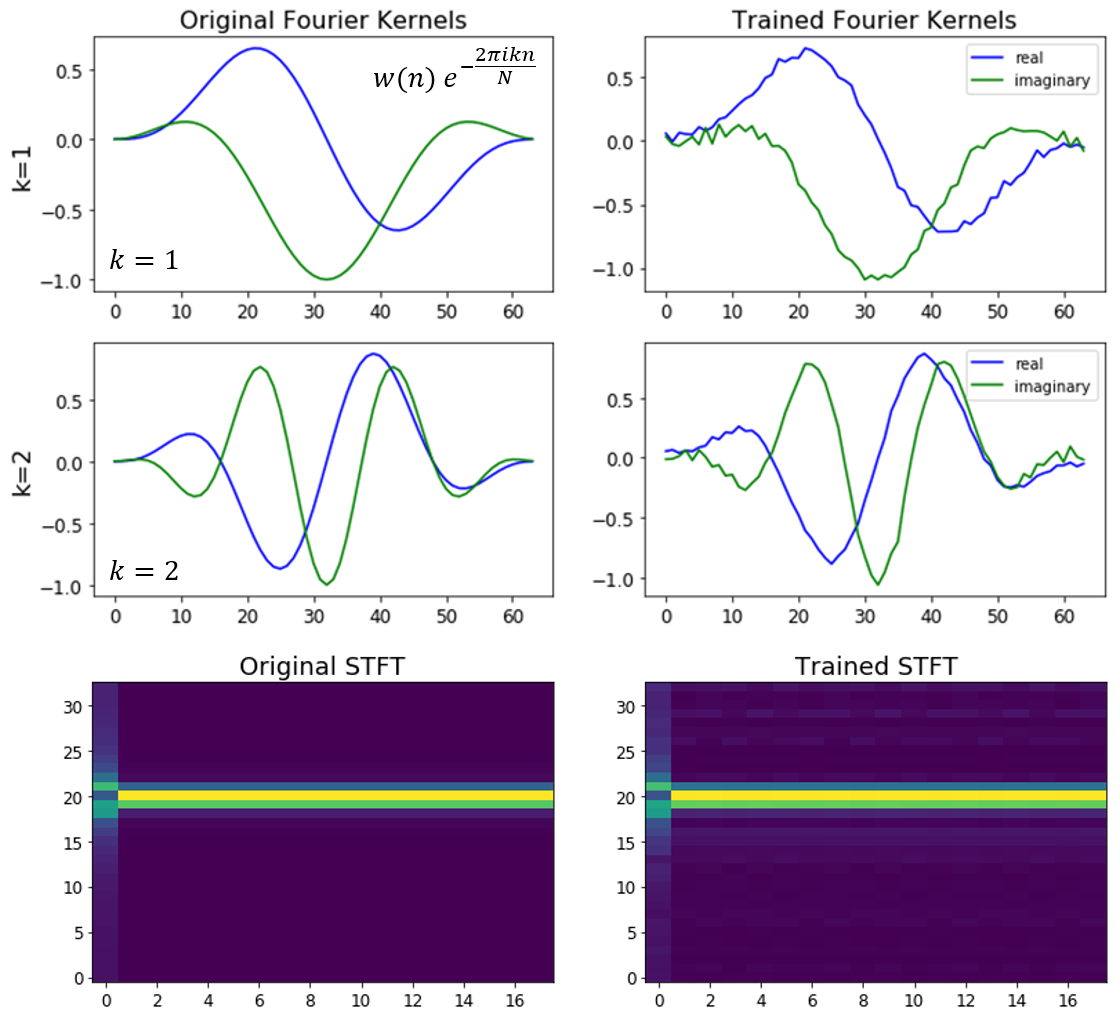}
{The first two rows show the Fourier kernels before and after training (only two of the kernels are shown here), and the third row shows the spectrograms resulting from the original and trained kernels.  \label{fig: Trained_STFT}}
\Figure(topskip=0pt, botskip=0pt, midskip=0pt)[width=80mm]{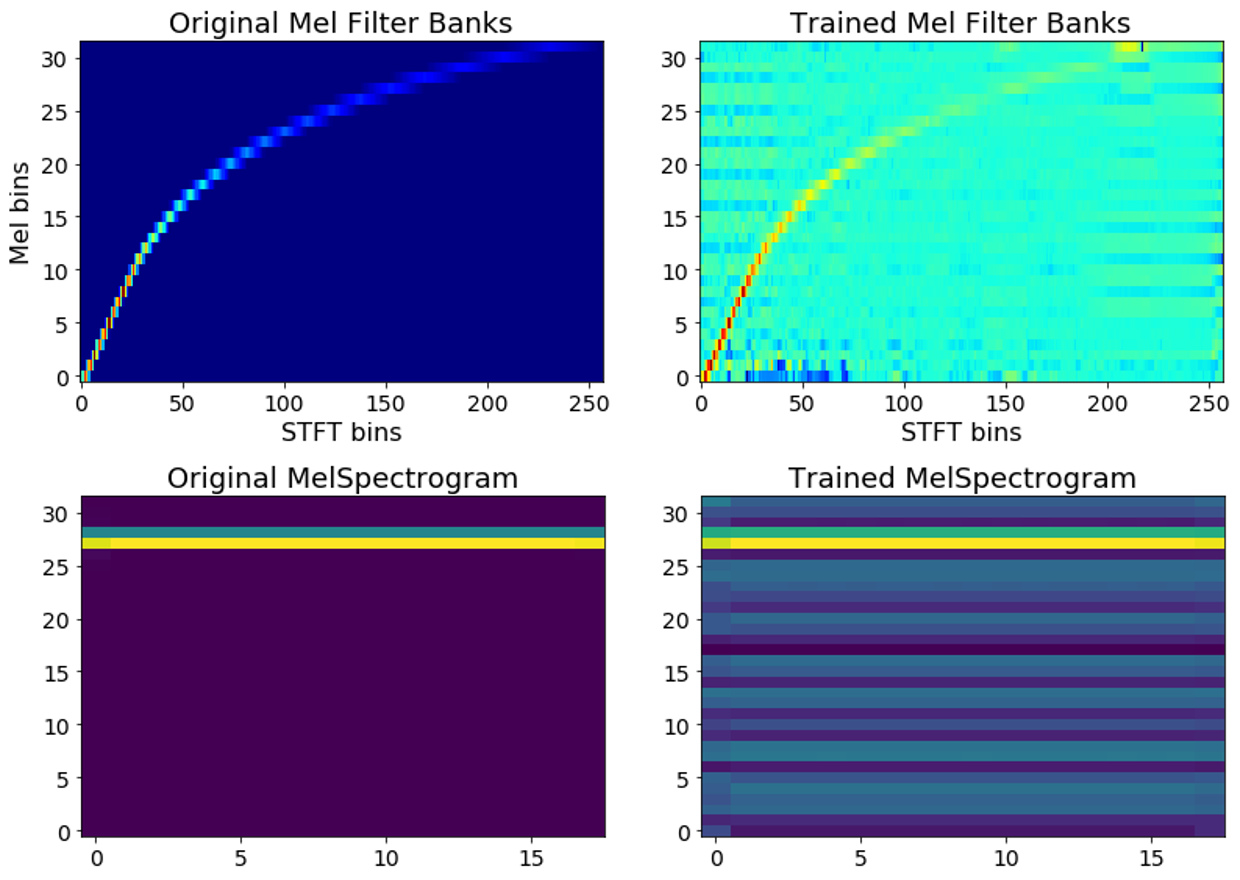}
{The first row shows the complete set of Mel filter banks before and after training. The resulting spectrograms are shown below.\label{fig: Trained_Mel}}
\Figure(topskip=0pt, botskip=0pt, midskip=0pt)[width=80mm]{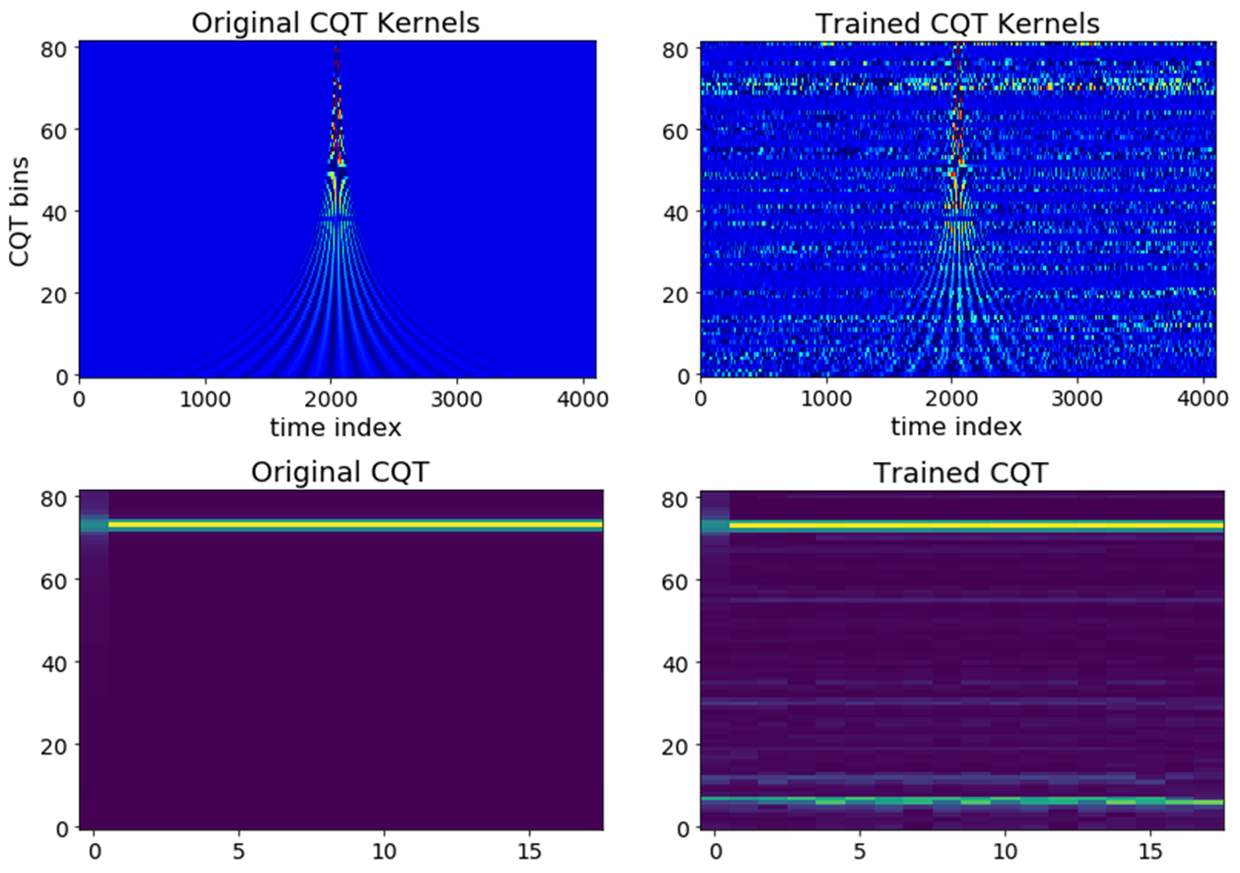}
{The first row shows the complete set of CQT kernels before and after training. Their resulting spectrograms are shown below. \label{fig: Trained_CQT}}

\section{Conclusion}
We propose an on-the-fly GPU based library, nnAudio, for spectrogram extraction. Different time domain to frequency domain transformation algorithms such as short-time Fourier transform, Mel spectrograms, and constant-Q transform have been implemented in \nbh{PyTorch}, an open source machine learning library. We leverage the CUDA integration of \nbh{PyTorch} the enable GPU based audio processing. GPU audio processing reduces time it takes to convert 1,770 waveforms to spectrograms conversion from $10.6$ seconds to only $0.001$ seconds for the Short-Time Fourier Transform (STFT); from $18.3$ seconds to $0.015$ seconds for Mel spectrogram; and from $103.4$ seconds to $0.258$ seconds for constant-Q Transform (CQT). These experiments were performed on a DGX station with CPU: Intel(R) Xeon(R) CPU E5-2698 v4 @ 2.20GHz and 4 Tesla v100 32Gb GPUs. (Only one GPU was utilized for the experiments.) Although it takes some time (around 5 seconds) to initialize the transformation layer (transferring Fourier kernels from RAM to GPU memory), once everything is ready on the GPU memory, the processing time is in the micro-seconds level for a single spectrogram, making the initialization time negligible in the context of training a neural network. 

As a second contribution of this paper, we further improve the existing CQT algorithms by proposing a method that does not require the calculation of the frequency domain CQT kernels. We directly apply the time domain CQT kernels on the original waveforms (in the time domain). This eliminates the needs to convert the waveforms to the frequency domain with a STFT. As a result, we need less computational steps, and hence have a faster CQT computation. The CQT computation speed for converting 1,770 waveforms to spectrograms is reduced drastically: from $0.258$ to only $0.001$ with our proposed algorithm when using GPU.

Finally, by implementing our proposed transformation algorithms as neural networks, the resulting the Fourier kernels, Mel filter banks, and even CQT kernels are trainable and finetunable. A small experiment confirms that trainable kernels can result in a better final model on a frequency prediction task compared to non-trainable kernels. 

We combined all of the discussed algorithms into a user-friendly PyPI package called nnAudio, so that it is easy for other researchers to use our proposed GPU audio processing tool\footnote{\url{https://github.com/KinWaiCheuk/nnAudio}.}

\vfill\pagebreak

\bibliography{citation.bib}
\bibliographystyle{IEEEtranN} 

\vfill\pagebreak

\begin{IEEEbiography}[{\includegraphics[width=1in,height=1.25in,clip,keepaspectratio]{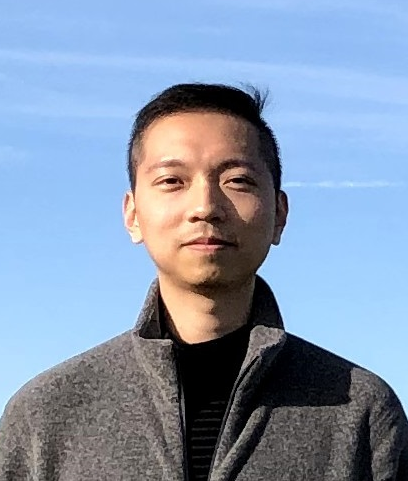}}]{Kin Wai Cheuk} (S'19) is a Ph.D student at Singapore University of Technology and Design (SUTD), supervised by Dr. Herremans and affiliated with the Institute of High Performance Computing at the Agency for Science Technology and Research, A*STAR. He received a B.S. degree in physics and M.Phil in mechanical engineering from the University of Hong Kong in 2014 and 2016 respectively. After graduation, he worked for an IT solution company and had a growing interest in machine learning during this period. In 2018, he joined Dr. Herremans' research group and started working on music related machine learning projects. He started his Ph.D in 2019 at SUTD. He is interested in digital signal processing and music transcription using machine learning. He received the Singapore International Graduate Award (SINGA) for his Ph.D degree.
\end{IEEEbiography}

\begin{IEEEbiography}[{\includegraphics[width=1in,height=1.25in,clip,keepaspectratio]{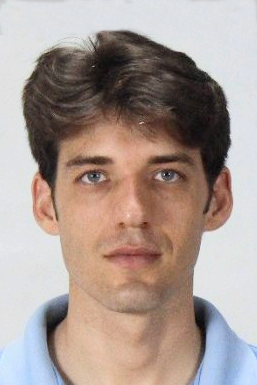}}]{Hans Anderson} is director of Blue Mangoo Software, a company based in Ha Noi that develops musical audio applications for mobile phones and tablets. He is also a freelance audio signal processing consultant. He completed his doctoral studies at the Singapore University of Technology and Design in 2018, and his master's degree in computational mathematics at the University of Minnesota in 2007. He loves designing signal processing algorithms for practical applications and commercial products, with a focus on improving the subjective sound quality of widely-used methods. Recently, he is working on eliminating aliasing noise from nonlinear audio effects such as saturation, compression, and amplifier models.
\end{IEEEbiography}

\begin{IEEEbiography}[{\includegraphics[width=1in,height=1.25in,clip,keepaspectratio]{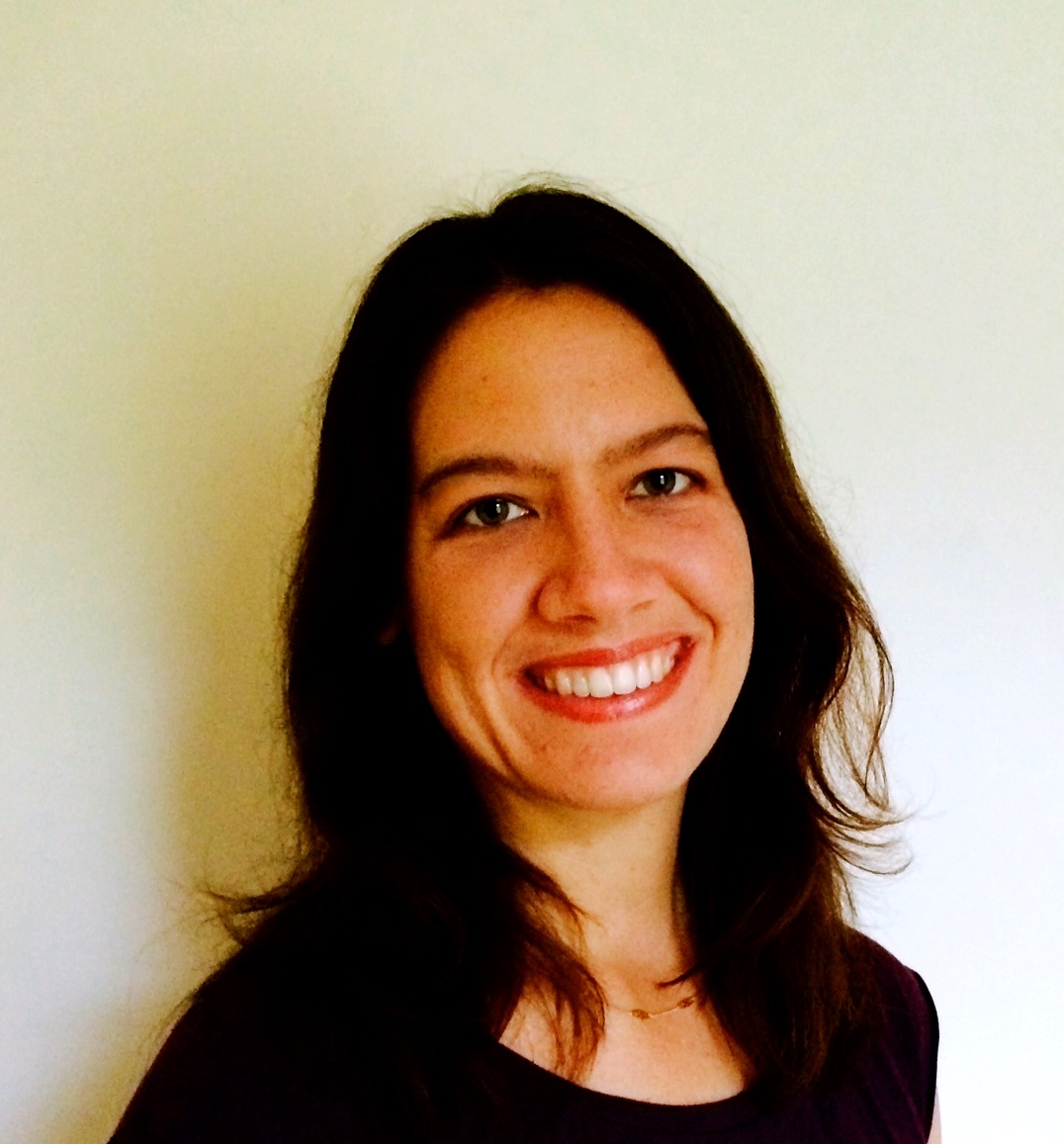}}]{Kat Agres} (M'17) is an Assistant Professor at the National University of Singapore (NUS), and a Research Scientist and Principal Investigator of the Music Cognition group at the Institute of High Performance Computing (IHPC), which is housed within the Agency for Science, Technology and Research (A*STAR). She received her PhD in Psychology with a graduate minor in Cognitive Science from Cornell University in 2013. She also holds a bachelor's degree in Cognitive Psychology and Cello Performance from Carnegie Mellon University, and has received numerous grants to support her research, including a Fellowship from the National Institute of Health (NIH), and a training fellowship from the National Institute of Mental (NIMH). Before moving to Singapore, she completed two postdoctoral research positions in the School of Electronic Engineering and Computer Science at Queen Mary University of London (QMUL). Her research explores a wide range of topics, including music technology for healthcare and well-being, music perception and cognition, computational creativity, information theoretic modelling of learning and memory, statistical learning, and the evaluation of creativity in humans and artificial systems.
\end{IEEEbiography}
\vfill\null

\begin{IEEEbiography}[{\includegraphics[width=1in,height=1.25in,clip,keepaspectratio]{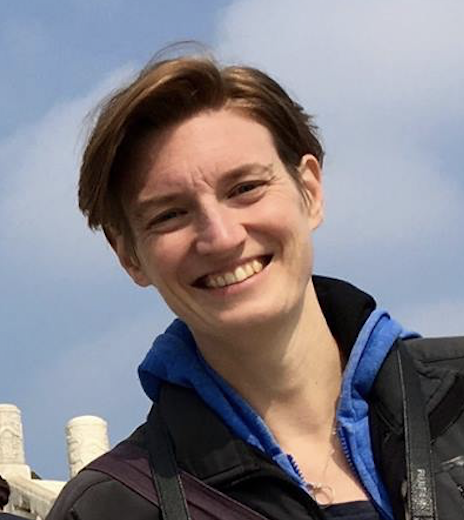}}]{Dorien Herremans} (M'12, SM'17) is an Assistant Professor at Singapore University of Technology and Design, with a joint appointment at the Institute of High Performance Computing at the Agency for Science Technology and Research, A*STAR. In 2015, she was awarded the individual Marie-Curie Fellowship for Experienced Researchers, and worked at the Centre for Digital Music, Queen Mary University of London. Prof. Herremans received her PhD in Applied Economics from the University of Antwerp. After graduating as a commercial engineer in management information systems at the University of Antwerp in 2005, she worked as a Drupal consultant and was an IT lecturer at Les Roches University in Bluche, Switzerland. Prof. Herremans' research focuses on the intersection of machine learning/optimization and digital music/audio. She is a Senior Member of the IEEE and co-organizer of the First International
Workshop ton Deep Learning and Music as part of IJCNN, as well as guest editor for Springer's Neural Computing and Applications.
\end{IEEEbiography}
\vfill\null
\EOD

\end{document}